\documentclass[Afour,sageh,times]{sagej}

\usepackage{moreverb}
\usepackage[utf8]{inputenc}

\newcommand\BibTeX{{\rmfamily B\kern-.05em \textsc{i\kern-.025em b}\kern-.08em
T\kern-.1667em\lower.7ex\hbox{E}\kern-.125emX}}

\def\volumeyear{2023}

\usepackage{mathtools}

\usepackage{url}
\usepackage{comment}
\usepackage[final,bookmarks=false]{hyperref}
\usepackage{cleveref}
\Crefname{section}{Section}{Sections}
\Crefname{figure}{Figure}{Figures}
\Crefname{table}{Table}{Tables}
\crefname{lstlisting}{listing}{listings}
\Crefname{lstlisting}{Listing}{Listings}

\usepackage{xspace}
\usepackage{multirow}
\usepackage{float}
\usepackage{array}
\usepackage{tikz}
\usetikzlibrary{mindmap,shadows,patterns,calc}
\usepackage[framemethod=tikz]{mdframed}
\definecolor{usethiscolorhere}{rgb}{0.86666,0.78431,0.78431}

\usepackage{amssymb}
\usepackage{booktabs}
\usepackage{pifont}
\usepackage{multirow}
\usepackage{paralist}

\usepackage[ruled,norelsize]{algorithm2e}
\makeatletter
\newcommand{\removelatexerror}{\let\@latex@error\@gobble}
\makeatother

\usepackage{listings}
\usepackage{graphicx}
\usepackage{subfig}
\usepackage{caption}
\usepackage{wrapfig}
\usepackage{colortbl}
\usepackage{xcolor}
    \definecolor{ListingsKeywordColor}{rgb}{0,0,0.4}
    \definecolor{ListingsIdentifierColor}{rgb}{0,0.5,0}
    \definecolor{ListingsCommentColor}{rgb}{0.4,0.4,0.4}
    \definecolor{ListingsStringColor}{rgb}{0.6000,0.3333,0.7333}
    \definecolor{ListingsRuleSepColor}{rgb}{0,0,0}
    \definecolor{ListingsEmphColor}{rgb}{0,0.6667,0.6667}
    \definecolor{ListingsBreakSymbolColor}{rgb}{0.780,0.082,0.522}
    \definecolor{LinkColor}{rgb}{0,0,0.5}
    \definecolor{UnitColor}{rgb}{0,0,0}
    \definecolor{MathsVectorColor}{rgb}{0,0,0}
    \definecolor{MathsMatrixColor}{rgb}{0,0,0}
    \definecolor{MyGreen}{HTML}{228B22}
    \definecolor{MyBlue}{HTML}{0000FF}
    \colorlet{MatrixElementsLight}{gray!40!white}
    \colorlet{MatrixElementsDark}{gray!80}
    \colorlet{MyGreenLight}{MyGreen!40!white}
    \colorlet{MyGreenDark}{MyGreen!80}    
    \colorlet{MyBlueLight}{MyBlue!40!white}
    \colorlet{MyBlueDark}{MyBlue!80}    
    \colorlet{MyRedLight}{red!20!white}
    \colorlet{MyRedDark}{red!60}

\def\twosum{\texttt{twosum}}
\def\twoprod{\texttt{twoprod}}
\def\binary{\texttt{binary64}}
\def\fma{\texttt{fma}}
\def\blas{BLAS}
\def\exblas{ExBLAS}

\def\axpy{{\sc axpy}}

\def\dotp{{\sc dot}}
\def\exdotp{{\sc exdot}}

\def\CC{{ C\nolinebreak[4]\hspace{-.05em}\raisebox{.4ex}{\scriptsize\bf ++ }}}

\newcommand{\becomes}{:=}

\usepackage{amsmath}

\renewcommand{\dotp}{{\sc dot}\xspace}
\renewcommand{\axpy}{{\sc axpy}\xspace}
\newcommand{\axpyl}{{\sc axpy}(-like)\xspace}

\newcommand{\spmv}{{\sc SpMV}\xspace}

\newcommand{\bd}{e}

\begin{document}

\title{General framework for re-assuring numerical reliability in parallel Krylov solvers: A case of BiCGStab methods}

\author{Roman Iakymchuk\affilnum{1,2},
Jos\'e I. Aliaga\affilnum{3}
}

\affiliation{\affilnum{1}Ume\aa\ University, Sweden\\
\affilnum{2}Sorbonne University, France\\
\affilnum{3}Universitat Jaime I, Spain
}

\corrauth{Roman Iakymchuk, Ume\aa\ University, MIT-Huset, 901 87 Ume\aa, Sweden}
\email{riakymch@cs.umu.se}

\begin{abstract}
Parallel implementations of Krylov subspace methods often help to accelerate the procedure of finding an approximate solution of a linear system. 
However, such parallelization coupled with asynchronous and out-of-order execution often enlarge the non-associativity impact in floating-point operations. 
These problems are even amplified when communication-hiding pipelined algorithms are used to improve the parallelization of  Krylov subspace methods.
Introducing reproducibility in the implementations avoids these problems by getting more robust and correct solutions.    
This paper proposes a general framework for deriving reproducible and accurate variants of Krylov subspace methods. 
The proposed algorithmic strategies are reinforced by programmability suggestions to assure deterministic and accurate executions. 
The framework is illustrated on the preconditioned BiCGStab method and its pipelined modification, which in fact is a distinctive method from the Krylov subspace family, for the solution of non-symmetric linear systems with message-passing.
Finally, we verify the numerical behaviour of the two reproducible variants of BiCGStab on a set of matrices from the SuiteSparse Matrix Collection and a 3D Poisson's equation.
\end{abstract}

\maketitle

\section{Introduction}
Solving large and sparse linear systems of equations appears in many scientific applications spanning from circuit and device simulation, quantum physics, large-scale eigenvalue computations, 
 and up to all sorts of applications 
that include the discretization of 
partial differential equations (PDEs)~\cite{barrettemplates}.
In this case, Krylov subspace methods fulfill the roles of standard linear algebra solvers~\cite{Saa03}. The Conjugate Gradient (CG) method can be considered as a pioneer of such iterative solvers operating on symmetric and positive definite (SPD) systems. 
Other Krylov subspace methods have been proposed to find the solution of more general classes of non-symmetric and indefinite linear systems. 
These include the Generalized Minimal Residual method (GMRES)~\cite{gmres}, 
the Bi-Conjugate Gradient (BiCG) method~\cite{bicg}, 
the Conjugate Gradient Squared (CGS) method~\cite{cgs}, 
and the widely used BiCG stabilized (BiCGStab) method by Van der Vorst~\cite{vandervorst} as a smoother converging version of the above two. 
Moreover, preconditioning is usually incorporated in real implementations of these methods in order to accelerate the convergence of the methods and improve their numerical features.

One would expect that the results of the sequential and parallel implementations of Krylov subspace methods to be identical, for instance, in the number of iterations, the intermediate and final residuals, as well as the sought-after solution vector. 
However, in practice, this is not often 
the case due to different reduction trees -- the Message Passing Interface (MPI) libraries 
offer up to 14 different implementations for reduction --, 
data alignment, instructions used, etc. 
Each of these factors impacts the order of floating-point operations, which are commutative but not associative, and, therefore, violates reproducibility. 
We aim to ensure identical and accurate outputs of computations, including the residuals/ errors, as in our view this is a way to ensure {\em robustness} and {\em correctness} of iterative methods. 
In this case, the robustness and correctness have a threefold goal: {\em reproducibility}\footnote{Reproducibility is the ability to obtain a bit-wise identical and  accurate result for multiple executions on the same data in various parallel environments.} of the results with the {\em accuracy guarantee} as well as {\em sustainable (energy-efficient)} algorithmic solutions.

The implementation of Krylov subspace methods on massively parallel systems reveals their scalability problems.
Mainly, because the synchronization of global communications, especially the reductions, delays parallel executions.
The most common solution has been the developments of communication-avoiding methods and, also, the use of new MPI functions to hide the communications, overlapping their execution with the computation of iterative methods.
In~\cite{cools17}, the authors propose a general framework for deriving {\em pipelined Krylov subspace methods}, in which the recurrences are reformulated to make easier the parallelization.
Again, these changes impact on the robustness and correctness of the iterative methods. 

In general, Krylov subsbpace methods are built from three components: sparse-matrix vector multiplication $Ax$ ({\spmv}), \dotp product between two vectors $(x,y)$, and scaling a vector by a scalar with the following addition of two vectors $y:=\alpha x + y$ (\axpy). 
If a block data distribution is used, only \axpy is performed locally, while \spmv needs to gather the full operand vector, e.g. via the {\tt MPI\_Allgatherv()} collective, and \dotp product requires communication and computation, e.g. via the {\tt MPI\_Allreduce()} collective, among MPI processes. 
Although \spmv has the highest amount of floating-point operations (flops), at large scale \dotp product becomes the most time-consuming component of Krylov subspace methods due to the required global communication. 
This justifies the use of pipelined versions of Krylov subspace methods.

In this paper, we aim to re-ensure reproducibility of Krylov subspace methods in parallel environments. Our contributions are the following: \vspace*{-0.1cm}
\begin{itemize}
    \item we propose a {\em general framework for deriving reproducible Krylov subspace methods}. 
    We follow the bottom-up approach and ensure reproducibility of Krylov subspace methods via reproducibility of their components, including the global communication. 
    We build our reproducible solutions on the \exblas~\cite{Collange15Parco} approach and its lighter version.
    \item even when applying our reproducible solutions, we particularly stress the importance of arranging computations carefully, e.g. avoid possibly replacements by compilers of $a*b+c$ in the favor of fused multiply-add (\fma) operation or postponing divisions in case of data initialization (i.e. divide before use). For instance, we provide customized~\axpyl operations using \fma{}, which reduces round-offs to one or two per \axpyl operation. 
    We refer to the 30-year-old but still up-to-date guide ``What every computer scientist should know about floating-point arithmetic" by Goldberg~\cite{goldberg}.
    \item we verify the applicability and performance of the proposed methodology on the preconditioned BiCGStab (PBiCGStab) and the pipelined PBiCGStab method. 
    We derive two reproducible variants of each method and test them on a set of SuiteSparse matrices and a 3D Poisson's equation.
\end{itemize}

This paper is structured as follows. 
\Cref{sec:background} reviews several aspects of computer arithmetic as well as the \exblas\ approach. 
\Cref{sec:framework} proposes a general framework for constructing reproducible Krylov subspace methods.
\Cref{sec:alg} introduces the PBiCGStab and the pipelined PBiCGStab methods, describing their MPI implementation in detail. 
Later, we evaluate the two reproducible implementations of PBiCGStab and pipelined PBiCGStab in~\Cref{sec:results}. 
Finally, 
\Cref{sec:related:works} reviews related work, while 
\Cref{sec:conclusion} draws conclusions and outlines future directions.

\section{Background}
\label{sec:background}
At first, we briefly introduce the floating-point arithmetic that
consists in approximating real numbers by numbers that have a finite, fixed-precision representation. 
These are composed of a significand, an exponent, and a sign:
\begin{equation*}
x = \pm \underbrace{x_0 . x_1 \ldots x_{M-1}}_{mantissa} \times b^{e}, \,\, 0 \leq x_i \leq b-1, \,\, x_0 \neq 0,
\end{equation*}
where $b$ is the  basis ($2$ in our case), $M$ is the precision, and $e$ stands for the exponent that is bounded ($e_{\min} \leq e \leq e_{\max}$).

The IEEE 754 standard~\cite{IEEE7542008}, created in 1985 and then revised in 2008 and in 2019, has
led to a considerable enhancement in the reliability of numerical computations by
rigorously specifying the properties of floating-point arithmetic. This standard
is now adopted by most processors, thus leading to a much better
portability of numerical applications.
The standard specifies floating-point formats, which are often associated with precisions like {\em binary16}, {\em binary32}, and {\em binary64}, see~\Cref{tb:ieee754}.
Floating-point representation allows numbers to 
cover a wide 
\textit{dynamic range} that
is defined as the absolute ratio between the number with the largest 
magnitude and the number with the smallest non-zero magnitude in a set. For instance, \binary\ (double-precision) can represent 
positive numbers from $4.9\times10^{-324}$ to $1.8\times10^{308}$, so it covers a dynamic range of $3.7\times10^{631}$.
\begin{table*}[]
\caption{Parameters for three IEEE arithmetic precisions.}
\label{tb:ieee754}
\centering
\begin{tabular}{llllll}
\hline
Type & Size & Significand & Exponent & Rounding unit & Range \\
\hline\noalign{\vskip .5mm} 
half & 16 bits & 11 bits & 5 bits & $u = 2^{-11} \approx 4.88 \times 10^{-4}$ & $\approx 10^{\pm 5}$ \\
single & 32 bits & 24 bits & 8 bits & $u = 2^{-24} \approx 5.96 \times 10^{-8}$ & $\approx 10^{\pm 38}$\\
double & 64 bits & 53 bits & 11 bits & $u = 2^{-53} \approx 1.11 \times 10^{-16}$ & $\approx 10^{\pm 
308}$\\
\hline
\end{tabular}
\end{table*}

The IEEE 754 standard requires correctly rounded results for the basic arithmetic operations $(+, -, \times , /, \sqrt{~},$ {\tt fma}$)$. It 
means that 
they 
are performed as if the result was first computed with an infinite precision and then rounded 
to the floating-point format. The correct rounding criterion guarantees a unique, well-defined answer, ensuring bit-wise reproducibility for a single operation; but correct rounding alone is not necessary to achieve reproducibility.
Emerging attention to reproducibility strives to draw more careful attention to the problem by the computer arithmetic community. It has led to the inclusion of error-free 
transformations (EFTs) for addition and multiplication -- to return the exact outcome as the result and the error -- to assure numerical reproducibility of floating-point operations,     
into the revised version of the 754 standard in 2019. 
These mechanisms, once implemented in hardware, will simplify our reproducible algorithms -- like the ones used in the ExBLAS~\cite{Collange15Parco}, ReproBLAS~\cite{Demmel14OneRed}, OzBLAS~\cite{ozblas} libraries -- and boost their performance.

There are two approaches that enable the addition of floating-point numbers without incurring round-off errors or with reducing their impact. 
The main idea is to keep track of both the result and the error during the course of computations.
The first approach uses EFT to compute both the result and the rounding error, storing them in a floating-point expansion (FPE). 
This is an unevaluated sum of $p$ floating-point numbers, 
whose components are ordered in magnitude with minimal overlap to cover the whole range of exponents. 
Typically, FPE relies upon the use of the traditional EFT for addition that 
is \twosum~\cite{Knu69}  and for multiplication that is \twoprod~\cite{Ogita05accuratesum}. 
The code of these two operations are, respectively, shown in \Cref{alg:TwoSum} and \Cref{alg:TwoProd}.
The second approach projects the finite range of exponents of floating-point numbers into a long vector so called a long (fixed-point) accumulator and stores every bit there. 
For instance, Kulisch~\cite{Kulisch11} proposed to use a 4288-bit long accumulator for the exact \dotp product of two vectors composed of \binary\ numbers; such a large long accumulator is designed to cover all the severe cases without overflows in its highest digit.
\begin{minipage}[t]{0.495\textwidth}
 \removelatexerror
 \begin{algorithm}[H]
   \caption{Error-free transformation for the summation of two floating-point numbers.}
   \label[algorithm]{alg:TwoSum}
    \SetKwProg{Fn}{Function}{}{}
    \KwIn{$a,b$ are two floating-point numbers.}
    \KwOut{$r,s$ are the result and the error, resp.}
    \Fn{$[r, s]$ = \rm\texttt{twosum}\,($a, b$)} {
      $r \becomes a + b$\\
      $z \becomes r - a$\\
      $s \becomes (a - (r - z)) + (b - z)$
    }
 \end{algorithm}
\end{minipage}
\begin{minipage}[t]{0.495\textwidth}
 \removelatexerror
 \begin{algorithm}[H]
   \caption{Error-free transformation for the product of two floating-point numbers.}
   \label[algorithm]{alg:TwoProd}
    \SetKwProg{Fn}{Function}{}{}
    \KwIn{$a,b$ are two floating-point numbers.}
    \KwOut{$r,s$ are the result and the error, resp.}    
    \Fn{$[r, s]$ = \rm\texttt{twoprod}\,($a, b$)} {
      $r \becomes a * b$\\
      $s \becomes \rm fma(a, b, -r)$\\
    }
 \end{algorithm}
\end{minipage}
 

The ExBLAS project
~\cite{Iakymchuk15ExBLAS} 
 is an attempt to derive fast, accurate, and reproducible BLAS library
by constructing  a multi-level approach for these operations that are tailored for various modern architectures with their
complex multi-level memory structures.
On one side, this approach is aimed to be fast to ensure similar performance compared to the non-deterministic parallel
versions. On the other side, the approach is aimed to preserve every bit of information before the final rounding to 
the desired format to assure correct-rounding and, therefore, reproducibility.
Hence, ExBLAS combines together long accumulator and FPE into algorithmic solutions as well as efficiently tunes and implements them on various architectures, including conventional CPUs, Nvidia and AMD GPUs, and Intel Xeon Phi co-processors (for details we refer to~\cite{Collange15Parco}).
Thus, \exblas\ assures reproducibility through assuring correct-rounding.

The corner stone of ExBLAS is the reproducible parallel reduction, which is at the core of many BLAS routines.
The ExBLAS parallel reduction relies upon FPEs with 
the \twosum\ EFT
and long accumulators, so it is correctly rounded and reproducible. 
In practice, the latter is invoked only once per overall summation that results in the little overhead (less than $8$\,\%) on accumulating large vectors.
Our interest in this paper is the \dotp product of two vectors, which is a crucial fundamental \blas\ operation. 
The \exdotp\ algorithm is based on the reproducible parallel reduction and 
the \twoprod\ EFT: the algorithm accumulates the result and the error of \twoprod\ EFT to same FPEs and then follows the reduction scheme. 
We derive its distributed version with two FPEs underneath (one for the result and the other for the error) that are merged at the end of computations. 
These and the other routines -- such as matrix-vector product, triangular solve, and matrix-matrix multiplication -- are distributed in the ExBLAS library\footnote{ExBLAS repository: \url{https://github.com/riakymch/exblas}} .

\section{General framework for reproducible Krylov solvers}
\label{sec:framework}

This section provides the outline of a general framework for deriving a reproducible version of any traditional Krylov subspace method. 
The framework is based on two main concepts: 1) identifying the issues caused by parallelization and, hence, the  non-associativity of floating-point computations; 2) carefully mitigating these issues primarily with the help of computer arithmetic techniques as well as programming guidelines. 
The framework was implicitly used for the derivation of the reproducible variants of the Preconditioned Conjugate Gradient (PCG) method~\cite{iakymchuk20ijhpca,iakymchuk19jcam}. 


The framework considers the parallel platform to consist of $K$ processes (or MPI ranks), denoted as $P_1$, $P_2$, \ldots, $P_K$.
In this, the coefficient matrix $A$ is partitioned into $K$ blocks of rows ($A_1$, $A_2$, $\hdots$, $A_k$), 
where each $P_k$
stores one row-block with the $k$-th {\em distribution block} $A_k \in \mathbb{R}^{p_k \times n}$,
and $n=\sum_{k=1}^{K}p_k$.
Additionally, vectors are partitioned and distributed in the same way as $A$. For example, the residual vector $r$ is partitioned as $r_1$, $r_2$, $\hdots$, $r_K$ and $r_k$ is stored in $P_k$.
Besides, scalars are replicated on all $K$ processes.

\subsection{Identifying sources of non-reproducibility}~
The first step is to identify sources of non-associativy and, thus, non-reproduci\-bi\-li\-ty of the Krylov subspace methods in parallel environments. 
As it can verify in~\Cref{fig:krylov}, there are 
four common operations as well as message-passing communication patterns associated with them: sparse matrix-vector product (\spmv) and Allgatherv for gathering the vector\footnote{Certainly, there are better alternatives for banded or similar sparse matrices, but using {\tt MPI\_Allgatherv} is the simplified solution for nonstructured sparse matrices}, \dotp\ product with the Allreduce collective, scaling a vector with the following addition of two vectors (\axpy and \axpy-like), and the application of the preconditioner. 
Hence, we investigate each of 
them.
\begin{figure*}[tb]
\centering
\begin{minipage}{0.8\textwidth}
{\small
\begin{tabular}{|l|}
 \hline
 {\bf while} $(\tau > \tau_{\max})$              
 \\
\begin{minipage}{.95\textwidth}
\[
\begin{array}{l|l@{~}c@{~}l|l|l}
  \multicolumn{1}{l}{\mbox{\rm Step}}  &  \multicolumn{3}{l}{\mbox{\rm Operation}} & \mbox{\rm Kernel} & \mbox{\rm Communication} \\
\hline
S1: &   d &:=&A p                         & \mbox{\rm \spmv} & \mbox{\rm Allgatherv} \\
S2: &   \rho&:=&\beta/{<p,d>}       & \mbox{\rm \dotp~product} & \mbox{\rm Allreduce} \\
S3: &   x&:=&x+\rho p           & \mbox{\rm \axpy} & \mbox{\rm --} \\
S4: &   r&:=&r-\rho d           & \mbox{\rm \axpy} & \mbox{\rm --} \\
S5: &   y&:=&M^{-1} r           & \mbox{\rm Apply preconditioner} & \mbox{\rm depends} \\
S6: &   \alpha&:=&\beta           & \mbox{\rm scalar operation} & \mbox{\rm --} \\
S7: &   \beta&:=&y'*r           & \mbox{\rm \dotp~product} & \mbox{\rm Allreduce} \\
S8: &   \tau&:=& \sqrt{<r, r>}   & \mbox{\rm \dotp ~product + sqrt }& \mbox{\rm Allreduce} \\
S9: &   \alpha&:=&\beta / \alpha           & \mbox{\rm scalar operation} & \mbox{\rm --} \\
S10: &   p&:=&y+\alpha p           & \mbox{\rm \axpy-like} & \mbox{\rm --} 
\end{array}
\]
\end{minipage}
\\ {\bf end while}                                     
\\ \hline
\end{tabular}
}
\vspace{-3pt}
\caption{Preconditioned Conjugate Gradient method with annotated BLAS kernels and message-passing communication.}
\label{fig:krylov}
\end{minipage}
\end{figure*}

In general, associativity and reproducibility are not guaranteed when there is perturbation of floating-point operations in parallel execution. 
For instance, while invoking the {\tt MPI\_Allreduce()} collective operation cannot ensure the same result (its execution path) as it depends on the data, the network topology, and the underlying algorithmic implementation. 
Under these assumptions, \axpyl and \spmv are associativity-safe as they are performed locally on local slices of data. 
The application of preconditioner can also be considered safe, e.g. the Jacobi preconditioner, until all operations are reduction-free; more complex preconditioners will certainly raise an issue. 
Thus, the main issue of non-determinism emerges from parallel reductions
(steps $S2$, $S7$ and $S8$ in~\Cref{fig:krylov}).

\subsection{Re-assuring reproducibility}~
\label{sec:reassuring}
We construct our approach for reassuring re\-pro\-du\-ci\-bi\-li\-ty by primarily targeting \dotp products and parallel reductions.
Note that the non-deterministic implementation of the Krylov subspace method utilizes the \dotp routine from a BLAS library like Intel MKL followed by {\tt MPI\_Allreduce()}. Thus, we propose to refine this procedure into four steps:
\begin{itemize}
    \item exploit the \exblas\ and its lighter FPE-based versions to build reproducible and correctly-rounded \dotp\ product;
    \item extend the \exblas- and FPE-based \dotp\ products to distributed memory by employing {\tt MPI\_Allreduce()}. 
    This collective acts on either long accumulators or FPEs. 
    For the ExBLAS approach, we apply regular reduction, since the long accumulator is an array of long integers. 
    Note that we may need to carry an extra intermediate normalization after the reduction of $2*2^{K-1}$ long accumulators, where $K=64-52=12$ is the number of carry-safe bits per each digit of the long accumulator. 
    For the FPE approach, we define the MPI operation that is based on the \twosum\ EFT. Thus, at this point, the choice of the reduction algorithm underneath {\tt MPI\_Allreduce()} does not have an impact on the computations as every bit of information is stored;
	\item rounding to double: for long accumulators, we use the \exblas-native \texttt{Round()} routine. To guarantee correctly rounded results of the FPE-based computations, we employ the \texttt{NearSum} algorithm from~\cite{RuOgOi2008b}. It is worth mentioning that the rounding operation is performed locally and does not require any communication. In the previous versions of the code as in~\cite{iakymchuk22ppam}, we split the reduction into three steps: {\tt MPI\_Reduce()}, rounding, and {\tt MPI\_Bcast()}. However, this is negligible as we re-assure control of the reduction operation and, hence, eliminate the performance penalty of using two collectives with one extra synchronization.
\end{itemize}

It is evident that the results provided by \exblas\ \dotp are both correctly-rounded and reproducible. With the lightweight \dotp, we aim also to be generic and, hence, we provide the implementation that relies on FPEs of size eight with the early-exit technique. This way the working precision of the computations using FPEs is increased up to $8*52$ bits as mentioned in~\cite{Hida01} for the double-double arithmetic.  
Additionally, we add a check for the FPE-based implementations to cover a case when the condition number and/ or the dynamic range are too large and we cannot keep every bit of information. 
Then, the warning is thrown, containing also a suggestion to switch to the ExBLAS-based implementation. 
But, note that these lightweight implementations are designed for moderately conditioned problems or with moderate dynamic range in order be accurate, reproducible, but also high performing, since the ExBLAS version can be very resource demanding, specially on the small core count. 
To sum up, if the information about the problem is know in advance, it is worth pursuing the lightweight approach.



\subsection{Programmability effort}~
\label{sec:prog}
It is important to note that compiler optimization and especially the usage of the fused-multiply-and-add (\fma) instruction, which performs $a*b+c$ with the extended precision and the single rounding at the end, may lead to some non-deterministic results. 
For instance, in the \spmv computation, each MPI rank computes its dedicated part $d_k$ of the vector $d$ by multiplying a block of rows $A_k$ by the vector $p$.
Since the computations are carried locally and sequentially, they are deterministic and, thus, reproducible. 
However, some parts of the code like $a*b+c*d*e$ and $a+=b*c$ -- present in the original implementation of PBiCGStab -- may not always yield to the same result~\cite{reprofeltor}. 
This is due to the fact that, for performance reasons, the \CC language standard allows compilers to change the execution order of this type of operation. 
It also allows merging multiplications and summations with fused multiply-add (\fma) instructions. 
Hence, a compiler might translate $a*b+c*d$ to two multiplications $t1=a*b$ and $t2=c*d$, and a subsequent summation $t1+t2$; it might generate a single multiplication $t=c*d$
with a subsequent $\text{\fma}(a,b,t)$, 
which gives a slightly different result; or it may even compute $t=a*b$ first 
and then use the $\text{\fma}(c,d,t)$. 
Thus, we advise to instruct compilers to use \fma~explicitly via \texttt{std::fma} in \CC11, assuming the underlying architecture supports \fma.

Another important observation is to carefully perform divisions and initialization of data. 
For instance, the initialization of $b$ in the Krylov solvers is computed as $b=Ad$, being $d=\frac{1}{\sqrt{N}}(1,\dots,1)^T$.
In this case, we suggest to compute $b=Ad$ for $d=(1,\dots,1)^T$ first and then scale $b$ by $1/\sqrt{N}$, as we observed a slightly faster convergence (up to 7\,\%) for the Krylov solver.

\section{Reproducible BiCGStab}
\label{sec:alg}
The classic Biconjugate Gradient Stabilized method (BiCGStab)~\cite{vandervorst} was proposed as a fast and smoothly converging variant of the BiCG~\cite{bicg} and CGS~\cite{cgs} methods. 
We present here the preconditioned BiCGStab (PBiCGStab) and the pipelined preconditioned BiCGStab (p-PBiCGStab): their design and implementation with Message Passing Interface (MPI).

For both methods, we consider a linear system $Ax=b$, where the coefficient matrix 
$A \in \mathbb{R}^{n \times n}$ is sparse with $n_z$ nonzero entries;
$b \in \mathbb{R}^n$ is the right-hand side vector; and $x \in \mathbb{R}^n$ is the sought-after solution vector.

Additionally, and for simplicity, we integrate the Jacobi preconditioner~\cite{Saa03} in our implementations, which is composed of the diagonal elements of the matrix ($M = diag(A)$).
In consequence, the application of the preconditioner is conducted on a vector and requires an element-wise multiplication of two vectors.

\subsection{Message-passing Parallel PBiCGStab Implementation}
\label{sec:mpipbicgstab}


The algorithmic description of the classical iterative PBiCGStab is presented in~\Cref{fig:pbicgstab}.
The loop body consists of two \spmv ($S2$ and $S6$), 
two preconditioner applications ($S1$ and $S5$), 
five \dotp products ($S3$, $S7$, $S10$, and $S11$), 
six \axpyl operations ($S4$, $S8$, $S9$, and $S12$), 
and a few scalar operations~\cite{barrettemplates}.
\begin{figure*}[tb]
{\small
\begin{tabular}{|l|}
\hline
   Compute preconditioner for $A \rightarrow M$ 
\\ Set starting guess $x^{0}$ 
\\ Initiate $r^0:=b-Ax^{0}, p^0:=r^0,  \tau^0:=\parallel r^{0}\parallel_2, j := 0$     
\\ \hline
\vspace{-1ex}
\\ {\bf while} $(\tau^{j} > \tau_{\max})$ 
\\
\begin{minipage}{.95\textwidth}
\[
\begin{array}{l|l@{~}c@{~}l|l|l}
  \multicolumn{1}{l}{\mbox{\rm Step}}  &  \multicolumn{3}{l}{\mbox{\rm Operation}} & \mbox{\rm Kernel} 
& \mbox{\rm Comm}\\\hline
{S1}: &  \hat{p}^{j} & := &  M^{-1} p^{j} & \mbox{\rm Apply precond.} & \mbox{\rm --}
\\ {S2}: &  s^{j}& := & A\hat{p}^{j} & \mbox{\rm \spmv} & \mbox{\rm Allgatherv}
\\ {S3}: &  \alpha^{j}& := & < r^{0}, r^{j} >/<r^0, s^{j}> &  \mbox{\rm \dotp~product} & \mbox{\rm Allreduce}
\\ {S4}: &  q^{j} & := & r^{j}-\alpha^{j} s^{j} &  \mbox{\rm \axpy-like } & \mbox{\rm --}
\\ {S5}: &  \hat{q}^{j} & := &  M^{-1} q^{j} & \mbox{\rm Apply precond.} & \mbox{\rm --}
\\ {S6}: &  y^{j} & := &  A\hat{q}^{j} & \mbox{\rm \spmv } & \mbox{\rm Allgatherv}
\\ {S7}: &  \omega^{j} & := &   < q^{j}, y^{j} > / < y^{j}, y^{j} > & \mbox{\rm Two \dotp~products } & \mbox{\rm Allreduce}
\\ {S8}: &  x^{j+1} & := &  x^{j}+\alpha^{j} \hat{p}^{j} +\omega^{j} \hat{q}^{j} &  \mbox{\rm Two \axpy} & \mbox{\rm --}
\\ {S9}: &  r^{j+1} & := &  q^{j}-\omega^{j} y^{j} & \mbox{\rm  \axpy-like } & \mbox{\rm --}
\\ {S10}: &  \beta^{j} & := &   \frac{ < r^{0}, r^{j+1} >}{< r^{0}, r^{j} > } * \frac{ \alpha^{j}}{ \omega^{j} }  & \mbox{\rm \dotp product } & \mbox{\rm Allreduce}
\\ {S11}: &  \tau^{j+1} & := &  \parallel r^{j+1}\parallel_2  & \mbox{\rm \dotp~product + sqrt} & \mbox{\rm Allreduce}
\\ {S12}: &  p^{j+1} & := &   r^{j+1} + \beta^{j} ( p^{j} - \omega^{j} s^{j} )  & \mbox{\rm Two \axpy-like} & \mbox{\rm --}
\end{array}
\]
\end{minipage}
\\ {\bf end while}
\\ \hline
\end{tabular}
}
\vspace{-3pt}
\caption{
Formulation of the PBiCGStab solver annotated with computational kernels and communication. The threshold
$\tau_{\max}$ is an upper bound on the relative residual for the computed approximation to the solution.
In the notation, 
$<\cdot,\cdot>$ computes the \dotp\ (inner) product of its vector arguments.}
\label{fig:pbicgstab}
\end{figure*}


As described in~\Cref{sec:framework}, the framework includes a reproducible implementation of the most common operations in a parallel implementation of a Krylov subspace method. 
Therefore, we next perform a communication and computation analysis of a  message-passing implementation of the PBiCGStab solver. 
From there, we derive the reproducible version by following the guide from~\Cref{sec:framework}.

For clarity, hereafter we will drop the superindices that denote the iteration count in the variable names. 
Thus, for example, $x^{j}$ becomes $x$, where the latter 
stands for the storage space employed to keep the sequence of approximations $x^{0},x^{1},x^{2},\ldots$ computed during the iterative process.
Taking into account these previous considerations, we analyze the different computational kernels ($S1$--$S12$) that compose the loop body of a single PBiCGStab iteration in~\Cref{fig:pbicgstab}.

\paragraph*{Sparse matrix-vector product ($S2$, $S6$):}
This kernel needs as input operands: the coefficient matrix $A$, which is distributed by blocks of rows, and the corresponding vector ($\hat{p}$ or $\hat{q}$), which is partitioned and distributed using the same partitioning as $A$. 
For simplicity, we just explain below how $S2$ is computed. 

Prior to computing this kernel, we need to obtain a replicated copy of  the distributed vector $\hat{p}$ in all processes, denoted as $\hat{p} \rightarrow \bd$; vector $\bd$ is the only array that is replicated in all processes.
We can recognize here a communication stage, but, after that, 
each process can then compute its local piece of the output vector $s$ concurrently: $P_k: s_k = A_k \, \bd$.
This kernel thus requires assembling the distributed pieces of the vector $\hat{p}$ into a single vector $\bd$ that is replicated in all
processes (in MPI, for example via \texttt{MPI\_Allgatherv()}). 

The computation can then proceed in parallel, yielding the vector result $s$ in the expected
distributed state with no further communication involved. At the end, each MPI process owns the corresponding piece of the computed vector.
To ensure the reproducibility of this computation, the local \dotp product between the sparse rows of $A_k$ and $\bd$ is based on \fma\ as outlined in~\ref{sec:prog}.


\paragraph*{\dotp\ products ($S3$, $S7$, $S10$, $S11$):}
The next kernel in the loop body is the \dotp product in the step $S3$ between the distributed vectors $r^0$ and $s$. 
Here, each process can compute concurrently a partial result $P_k: \rho_k = <r^0_k,s_k>$ 
and when all processes have finished this partial computation, these intermediate values have to be reduced into a globally-replicated scalar 
$\alpha := \sigma/(\rho_1+\rho_2+\cdots+\rho_{K})$.
We can apply the same idea to the \dotp products in the steps $S7$, $S10$ and $S11$, yielding a total of five process synchronizations (in MPI, via {\tt MPI\_Allreduce()}) since 
all scalars are globally-replicated.
But, the number of synchronization can be reduced to four, considering that communications in S10 and S11 can be merged in a single {\tt MPI\_Allreduce()}.

The easiest solution to compute $\rho_k$ is to call to the \dotp routine from the Intel MKL or similar libraries, however this will not guarantee reproducibility even when \fma\ are used internally. 
Thus, we enforce reproducibility by applying our two ExBLAS-based strategies, following the guideline as in~\Cref{sec:reassuring}.

\paragraph*{\axpyl vector updates ($S4$, $S8$, $S9$, $S12$):}
The next kernel is the \axpy-like~kernel in the step $S4$, which involves 
the distributed vectors $q, r, s$ and the globally-replicated scalar $\alpha$. 
The operations in the steps $S8$, $S9$, and $S12$ follow the same idea because all scalars are globally-replicated. 
In this type of kernels, all processes can perform their local parts of the computation
to obtain the result without any communication: $P_k: q_k = r_k - \alpha \, s_k$.

While \axpy ($y = \alpha x + y$) can directly rely on the MKL library routine, \axpy-like ($z = \alpha y + x$) requires, at least, two routines in order to be implemented ({\sc scal}/ {\sc copy} + \axpy). 
Looking for a robust and correct solution, the use of MKL routines is a bad alternative since each one introduces a rounding error.
Additionally, this alternative is more expensive because some vector must be traversed more than once.
Instead, we propose to rely on \fma\ that computes each element of the solution of both \axpy and \axpy-like with a single rounding and only one pass through the vectors. 
Therefore, in the reproducible versions we provide our own implementations for \spmv, \axpy, and \axpy-like (do not rely on any external BLAS libraries) and, hence, have the overall control of computations, assuring their correct rounding and reproducibility.

\paragraph*{ Application of the preconditioner ($S1$, $S5$):}
The kernel in the step S1 consists of applying the Jacobi preconditioner $M$, 
scaling the vector $p$ by the diagonal of the matrix. 
Therefore, it can be executed in parallel by all processes because each of them stores a different set of the diagonal elements (those related with the piece of the matrix that it stores) and the corresponding set of the vector elements: $P_k: \hat{p}_k = M^{-1}_k p_k.$
The same procedure can be applied on the step $S5$ to scale the vector $q$, resulting in $\hat{q}$.

There is no routine in the MKL library to implement the element-wise product of two vector, therefore, an ad-hoc implementation has to be done.
Reproducibility is ensured if this code is based on \fma\ and the order of operations is deterministic as mentioned in~\Cref{sec:prog}.

\subsection{Message-passing Parallel p-PBiCGStab Implementation}~
\label{sec:mpippbicgstab}

The authors in~\cite{cools17} propose two main steps for deriving the pipelined version of a Krylov subspace method:
\begin{itemize}
    \item {\em Communication-avoiding}. 
    In which the number of global communications is reduced, 
    rearranging the original recurrences.
    Usually more terms appear in the new recurrences and, therefore, there are more vector operations.
    \item {\em Hiding communications}.
    Since global communications are the most time-consuming component of Krylov subspace methods at large scale, the alternative to reduce their impact on the performance of  parallel implementations is their simultaneous execution (overlapping) with \spmv.
    This technique is implemented using non-blocking collectives, such as \texttt{MPI\_Iallreduce()}, which require the use of \texttt{MPI\_Wait()} to check when the communication is complete.
\end{itemize}

The algorithmic description of the pipelined preconditioned BiCGStab (p-PBiCGStab) is presented in~\Cref{fig:ppbicgstab}.
The loop body consists of two \spmv\ ($S10$ and $S18$), 
two preconditioner applications ($S9$ and $S17$),
six \dotp\ products ($S8 \cup S11$ and $S16\cup S19$),
eighteen \axpy/ \axpy-like operations ($S1$-$S7$ and $S12$-$S15$),
and a few scalar operations~\cite{cools17}. It is worth mentioning that the pipelined PBiCGStab may show different convergence behavior compared to the standard PBiCGStab due to the different way floating-point operations are performed and, thus, the round off errors are propagated and accumulated differently.

The analysis of the computational kernels of the algorithm is very similar to the described above for the parallelization of PBiCGStab in~\Cref{sec:mpipbicgstab}.
The only difference is how the \dotp products are implemented.

\paragraph*{\dotp\ products ($S8 \cup S11$, $S16\cup S19$):}
Although, there are six \dotp in~\Cref{fig:ppbicgstab}, only two global synchronizations are required because more than one reduction is complete in each step.
Therefore, before the synchronization is initiated, the partial result related to the corresponding reductions has to be computed locally in each process.
Then, obtained values are stored in local vectors which are used to compute the global values using collectives.  
The overlapping requires the use of non-blocking collectives which decompose each reduction in two steps: 
the first step ($S8$ and $S16$) properly executes \texttt{MPI\_Iallreduce()} starting the global communication, which continues while other steps are performed, e.g., $S9$ and $S10$.
When the global values have to be used, the second step has to be done, calling \texttt{MPI\_Wait()}, since execution can only continue if the global communication is completed. We follow here the `golden rule' of the non-blocking communication -- start as soon as the data are available and wait right before they are needed.

\begin{figure*}[tb]
{\small
\begin{tabular}{|l|}
\hline
   Compute preconditioner for $A \rightarrow M$ 
\\ Set starting guess $x^{0}$ 
\\ Initiate $r^0:=b-Ax^{0}, \hat{r}^0:=M^{-1}r^{0}, w^0:=A\hat{r}^{0}, \hat{w}^0:=M^{-1}w^{0}, t^0:=A\hat{w}^{0}, \alpha^0:=<r^{0},r^{0}> / <r^{0},w^{0}> , \beta^{-1}:=0, j := 0$    
\\ \hline
\vspace{-1ex}
\\ {\bf while} $(\tau^{j} > \tau_{\max})$ 
\\
\begin{minipage}{.95\textwidth}
\[
\begin{array}{l|l@{~}c@{~}l|l|l}
  \multicolumn{1}{l}{\mbox{\rm Step}}  &  \multicolumn{3}{l}{\mbox{\rm Operation}} & \mbox{\rm Kernel} 
& \mbox{\rm Comm}\\\hline
{S1}: &  \hat{p}^{j} & := &  \hat{r}^{j}+\beta^{j-1}(\hat{p}^{j-1}-\omega^{j-1}\hat{s}^{j-1}) & \mbox{\rm Two \axpy-like} & \mbox{\rm --}
\\ {S2}: &  s^{j} & := &  w^{j}+\beta^{j-1}(s^{j-1}-\omega^{j-1}z^{j-1}) & \mbox{\rm Two \axpy-like} & \mbox{\rm --}
\\ {S3}: &  \hat{s}^{j} & := &  \hat{w}^{j}+\beta^{j-1}(\hat{s}^{j-1}-\omega^{j-1}\hat{z}^{j-1}) & \mbox{\rm Two \axpy-like} & \mbox{\rm --}
\\ {S4}: &  z^{j} & := &  t^{j}+\beta^{j-1}(z^{j-1}-\omega^{j-1}v^{j-1}) & \mbox{\rm Two \axpy-like} & \mbox{\rm --}
\\ {S5}: &  q^{j} & := &  r^{j}-\alpha^{j}s^{j} & \mbox{\rm \axpy-like} & \mbox{\rm --}
\\ {S6}: &  \hat{q}^{j} & := &  \hat{r}^{j}-\alpha^{j}\hat{s}^{j} & \mbox{\rm \axpy-like} & \mbox{\rm --}
\\ {S7}: &  y^{j} & := &  w^{j}-\alpha^{j}z^{j} & \mbox{\rm \axpy-like} & \mbox{\rm --}
\\ {S8}: & ~ & ~ & < q^{j}, y^{j} > , <y^{j}, y^{j}> &  \mbox{\rm Two \dotp~products} & \mbox{\rm Iallreduce}
\\ {S9}: &  \hat{z}^{j} & := &  M^{-1} z^{j} & \mbox{\rm Apply precond.} & \mbox{\rm --}
\\ {S10}: &  v^{j}& := & A\hat{z}^{j} & \mbox{\rm \spmv} & \mbox{\rm Allgatherv}
\\ {S11}: &  \omega^{j} & := &   < q^{j}, y^{j} > / < y^{j}, y^{j} > & \mbox{\rm Two \dotp~products } & \mbox{\rm Wait}
\\ {S12}: &  x^{j+1} & := &  x^{j}+\alpha^{j} \hat{p}^{j} +\omega^{j} \hat{q}^{j} &  \mbox{\rm Two \axpy} & \mbox{\rm --}
\\ {S13}: &  r^{j+1} & := &  q^{j}-\omega^{j} y^{j} & \mbox{\rm  \axpy-like } & \mbox{\rm --}
\\ {S14}: &  \hat{r}^{j+1} & := &  \hat{q}^{j}-\omega^{j} (\hat{w}^{j}-\alpha^{j} \hat{z}^{j}) & \mbox{\rm Two \axpy-like } & \mbox{\rm --}
\\ {S15}: &  w^{j+1} & := &  y^{j}-\omega^{j} (t^{j}-\alpha^{j} v^{j}) & \mbox{\rm Two \axpy-like } & \mbox{\rm --}
\\ {S16}: & ~ & ~ & < r^{0}, r^{j+1} > , <r^{0}, w^{j+1}> &  \mbox{\rm Four \dotp~products} & \mbox{\rm Iallreduce}
\\ ~ & ~ & ~ & < r^{0}, s^{j} > , <r^{0}, z^{j}> & ~ & ~
\\ {S17}: &  \hat{w}^{j+1} & := &  M^{-1} w^{j+1} & \mbox{\rm Apply precond.} & \mbox{\rm --}
\\ {S18}: &  t^{j+1}& := & A\hat{w}^{j+1} & \mbox{\rm \spmv} & \mbox{\rm Allgatherv}
\\ {S19}: &  \beta^{j} & := &   \frac{ < r^{0}, r^{j+1} >}{< r^{0}, r^{j} > } * \frac{ \alpha^{j}}{ \omega^{j} }  & \mbox{\rm Four \dotp~products } & \mbox{\rm Wait}
\\ ~ &  \alpha^{j+1} & := &   \frac{ < r^{0}, r^{j+1} >}{< r^{0}, w^{j+1} > + \beta^{j} < r^{0}, s^{j} > - \beta^{j} \omega^{j} < r^{0}, z^{j} >}   & ~~ & ~
\end{array}
\]
\end{minipage}
\\ {\bf end while}
\\ \hline
\end{tabular}
}
\vspace{-3pt}
\caption{
Formulation of the pipelined PBiCGStab solver annotated with computational kernels and communication. The threshold
$\tau_{\max}$ is an upper bound on the relative residual for the computed approximation to the solution.
In the notation, 
$<\cdot,\cdot>$ computes the \dotp\ (inner) product of its vector arguments.}
\label{fig:ppbicgstab}
\end{figure*}

\section{Experimental Results}
\label{sec:results}
In this section, we report a variety of numerical experiments to examine the convergence, scalability, accuracy, and reproducibility of the original and two reproducible versions of PBiCGStab and p-PBiCGStab.  
In our experiments, we employed IEEE754 double-precision arithmetic and
conducted them on the SkyLake partition at Fraunhofer with a dual Intel Xeon Gold 6132 CPU @2.6\,GHz, 28 cores, and 192\,GB of memory. Nodes are connected with the 54\,Gbit/s FDR Infiniband.  

\subsection{Evaluation on the SuiteSparse matrices\,\,} 
We carried out tests on a range of different linear systems from the SuiteSparse matrix collection on a single SkyLake node using 1, 2, 4, 8, 16, and 28 (full) cores. 
\Cref{tab:ssonenode} lists a set of tested matrices with 
 the number of rows/ columns $N$ and the number of nonzeros $nnz$. We aim to show the reproducibility, accuracy, and performance of our algorithmic implementations on matrices with various loads, i.e. number of nonzeros, as well as complexities.  
 The right-hand side vector $b$ in the iterative solvers was always initialized to the product $Ad, d=\frac{1}{\sqrt{N}}(1,\dots,1)^T$, where N is the number of rows/ columns of $A$. 
 However, in both \exblas- and FPE-based versions, marked as ReproPBiCGStab in the table, we computed $b=Ad, d=(1,\dots,1)^T$ and then scaled $b$ by $\frac{1}{\sqrt{N}}$. 
 In all implementations, iterations were started with the initial guess $x_0=0$. 
 The parameter that controls the convergence of the iterative process is $\|r^j\|_2/\|r^0\|_2\leq 10^{-6}$. 

\Cref{tab:ssonenode} reports the number of required iterations to reach the stopping criterion as well the final true residual for PBiCGStab and ReproPBiCGStab; the latter marks both \exblas- and FPE-based variants as they report identical results independently from the number of cores/ MPI processes used. We also report the initial residual ($\|r_0\|_2$) which can serve as an indicator in combination with the final true residual of how the convergence unfolds.
For the original version, we display the number of iterations on single and 16 cores as they differ.
Notably, the two reproducible variants show a tendency to deliver slightly better and more reliable accuracy of the approximate result (the final true residual) and/ or converge faster. For instance, the reproducible variants require significantly less iterations for the {orsreg\_1}, {rdb3200l}, {Transport}, {tmt\_unsym} matrices; for another six both are slightly faster and for only two are slightly slower. For a very large matrix as vas\_stokes\_2M, the standard non-deterministic PBiCGStab method was not able to converge while using 16 or 28 MPI processes.

\begin{table*}[tb]
\begin{center}
\resizebox{\textwidth}{!}{
\begin{tabular}{|l|l|r|r|r|r|r|r|r|r|}
   \hline
   Matrix  & Prec & $N$ & nnz & $\|r^0\|_2$ & \multicolumn{3}{c|}{PBiCGStab} & \multicolumn{2}{c|}{ReproPBiCGStab} \\  
   \cline{6-8}\cline{9-10}
   & & & & & iter1 & iter16 & $\|b-Ax^j\|_2$ & iter & $\|b-Ax^j\|_2$ \\\hline
   \rowcolor[gray]{.9}add32 & Jac & 4,960 & 19,848 & 8.00e-03 & 36 & 35 & 4.97e-09 & 35 & 7.12e-09\\ 
   af\_shell10 & Jac & 1,508,065 & 52,259,885 & 1.48e+05 & 9 & 9 & 2.18e-02 & 9 & 2.18e-02\\
   \rowcolor[gray]{.9}atmosmodd & Jac & 1,270,432 & 8,814,880 & 3.75e+03 & 221 & 256 & 2.68e-03 & 222 & 9.55e-04\\ 
   atmosmodj & Jac & 1,270,432 & 8,814,880 & 3.75e+03 & 227 & 227 & 3.46e-03 & 229 & 3.25e-03\\
   \rowcolor[gray]{.9}atmosmodl & Jac & 1,489,752 & 10,319,760 & 1.85e+04 & 133 & 133 & 1.80e-02 & 132 & 1.68e-02\\ 
   atmosmodm & Jac & 1,489,752 & 10,319,760 & 3.50e+05 & 77 & 77 & 2.43e-01 & 75 & 2.41e-01\\
   \rowcolor[gray]{.9}audikw\_1 & Jac & 943,695 & 77,651,847 & 1.58e+07 & 11 & 11 & 8.14e+00 & 11 & 8.04e+00\\ 
   bcsstk18 & Jac & 11,948 & 149,090 & 2.30e+09 & 7 & 7 & 7.51e+02 & 7 & 7.51e+02\\
   \rowcolor[gray]{.9}bcsstk26 & Jac & 1,922 & 30,336 & 6.16e+09 & 11 & 11 & 5.62e+03 & 11 & 5.62e+03\\
   bone010 & Jac & 986,703 & 47,851,783 & 8.55e+03 & 12 & 12 & 5.91e-03 & 12 & 5.91e-03\\
   \rowcolor[gray]{.9}boneS10 & Jac & 914,898 & 40,878,708 & 7.17e+03 & 12 & 12 & 3.92e-03 & 12 & 3.92e-03\\   
   Bump\_2911 & Jac & 2,911,419 & 127,729,899 & 1.91e+14 & 12 & 12 & 1.82e+08 & 12 & 1.82e+08\\
   \rowcolor[gray]{.9}cage14 & Jac & 1,505,785 & 27,130,349 & 1.00e+00 & 5 & 5 & 1.55e-07 & 5 & 1.55e-07\\
   cage15 & Jac & 5,154,859 & 99,199,551 & 1.00e+00 & 6 & 6 & 1.12e-07 & 6 & 1.12e-07\\
   \rowcolor[gray]{.9}circuit5M\_dc & Jac & 3,523,317 & 14,865,409 & 1.02e+04 & 5 & 5 & 6.52e-03 & 5 & 6.52e-03\\
   CurlCurl\_3 & Jac & 1,219,574 & 13,544,618 & 2.42e+10 & 17 & 17 & 2.14e+04 & 17 & 2.14e+04\\   
   \rowcolor[gray]{.9}CurlCurl\_4 & Jac & 2,380,515 & 26,515,867 & 2.10e+10 & 19 & 19 & 1.18e+04 & 19 & 1.18e+04\\   
   ecology1 & Jac & 1,000,000 & 4,996,000 & 1.96e+01 & 8 & 8 & 7.61e-06 & 8 & 9.66e-06\\ 
   \rowcolor[gray]{.9}ecology2 & Jac & 999,999 & 4,995,991 & 1.96e+01 & 8 & 8 & 7.38e-07 & 8 & 9.67e-07\\  
   Hardesty1 & Jac & 938,905 & 12,143,314 & 9.99e+00 & 17 & 20 & 9.28e-06 & 19 & 4.60e-06\\ 
   \rowcolor[gray]{.9}ML\_Geer & Jac & 1,504,002 & 110,686,677 & 4.89e+02 & 2886 & 3084 & 2.83e-04 & 3060 & 1.19e-04\\ 
   orsreg\_1 & Jac & 2,205 & 14,133 & 4.83e+00 & 225 & 243 & 4.18e-06 & 210 & 4.68e-06\\ 
   \rowcolor[gray]{.9}Queen\_4147 & Jac & 4,147,110 & 316,548,962 & 1.94e+14 & 52 & 51 & 3.61e+07 & 51 & 7.81e+07\\   
   rdb3200l & Jac & 3,200 & 18,880 & 9.96e+00 & 641 & 635 & 4.10e-06 & 583 & 3.17e-06\\ 
   \rowcolor[gray]{.9}s3dkq4m2 & Jac & 90,449 & 4,427,725 & 6.08e+02 & 23 & 23 & 7.27e-05 & 23 & 7.27e-05 \\
   tmt\_unsym & Jac & 917,825 & 4,584,801 & 6.45e-06 & 6489 & 6757 & 9.34e-12 & 5388 & 1.20e-11\\ 
   \rowcolor[gray]{.9}Transport & Jac & 1,602,111 & 23,487,281 & 2.45e-02 & 561 & 583 & 2.12e-08 & 557 & 1.74e-08\\ 
   vas\_stokes\_2M & Jac & 2,146,677 & 65,129,037 & 4.19e-01 & 6411 & - & 3.34e-07 & 6664 & 3.49e-07\\
   \hline
\end{tabular}
}
\end{center}
\caption {\label{tab:ssonenode} 
Convergence of the {\bf PBiCGStab and its reproducible versions} (ReproPBiCGStab, identical results reported for both) on a set of the SuiteSparse matrices. The initial guess is $x^0 = 0$. The number of iterations required to reach the tolerance of $10^{-6}$ on the scaled residual, i.e. $\|r^j\|_2/\|r^0\|_2$, is reported along with the corresponding true residual $\|b-Ax^j\|_2$.}
\end{table*}

\Cref{tab:ssonenode-pipe-iallreduce-fma} shows similar results but for the pipelined PBiCGStab and its reproducible variants. The tendency of reproducible variants to converge faster and to deliver more reliable accuracy is preserved. For instance, both variants are significantly faster for the {orsreg\_1}, {tmt\_unsym}, and vas\_stokes\_2M matrices; for another five matrices both are slightly faster, while for four are slightly slower. With the pipelined PBiCGStab, we were able to converge to the approximate solution of vas\_stokes\_2M under the tolerance of $10^{-6}$. When the required tolerance is increased to, e.g., $10^{-8}$ or higher, both the standard and, especially, the pipelined methods may not converge for vas\_stokes\_2M, ML\_Geer, and tmt\_unsym. We leave this as a future work and foresee to investigate this correlation between the requested accuracy and the abilities of the solvers, potentially employing some healing techniques like residual replacement.

\begin{figure*}[th!]
\begin{center}
\begin{tabular}{cc}
\begin{minipage}[t]{0.48\textwidth}
\includegraphics[width=\textwidth,height=6.2cm]{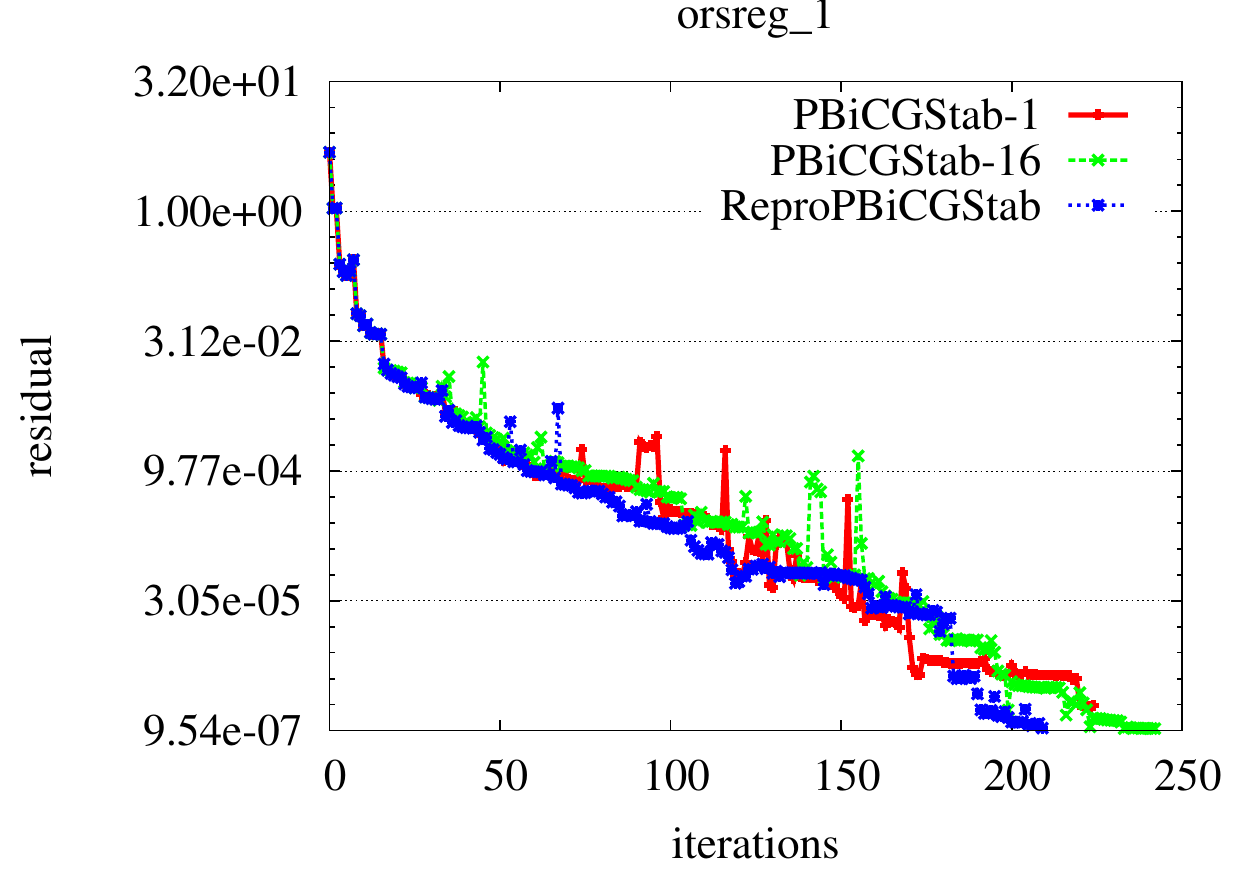}
\end{minipage}
&
\begin{minipage}[t]{0.48\textwidth}
\includegraphics[width=\textwidth,height=6.2cm]{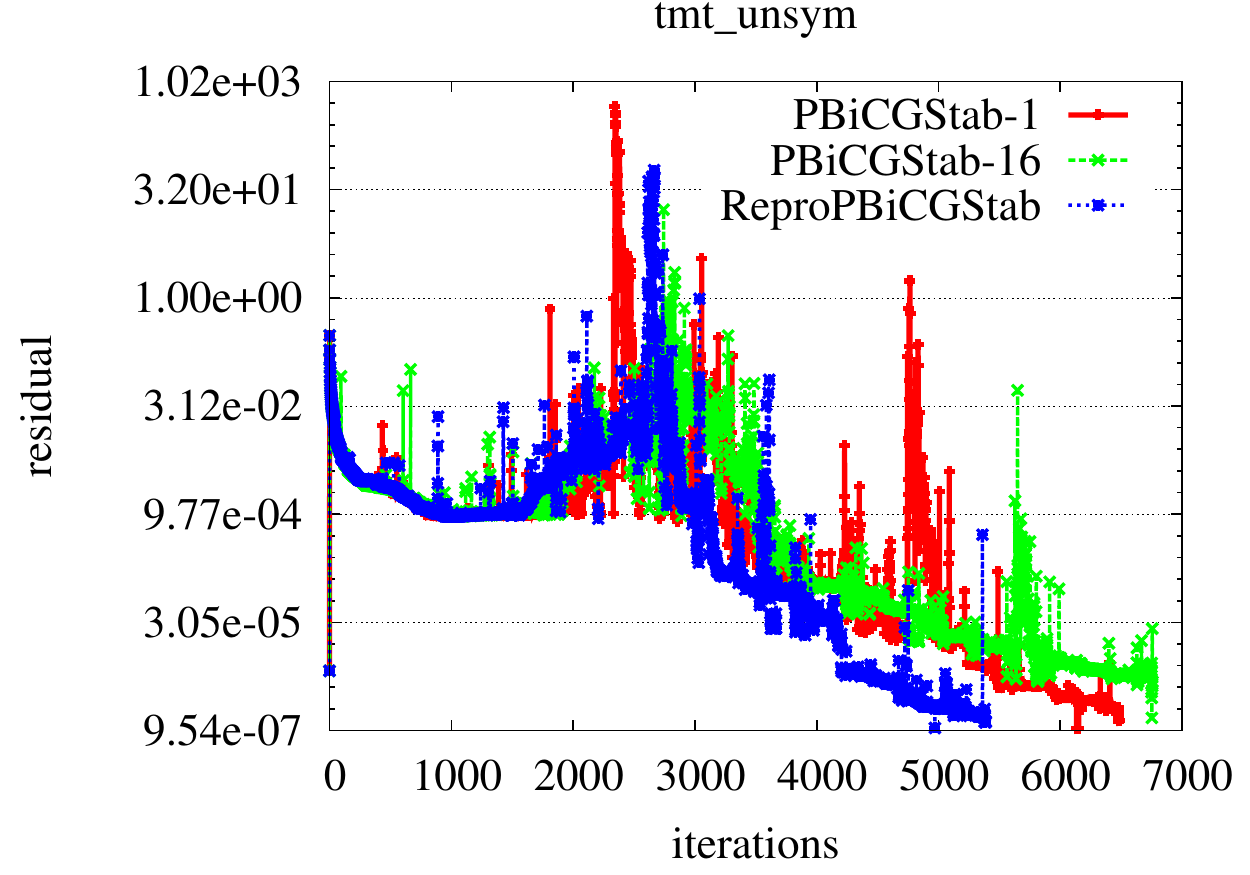}
\end{minipage}
\\
\begin{minipage}[t]{0.48\textwidth}
\includegraphics[width=\textwidth,height=6.2cm]{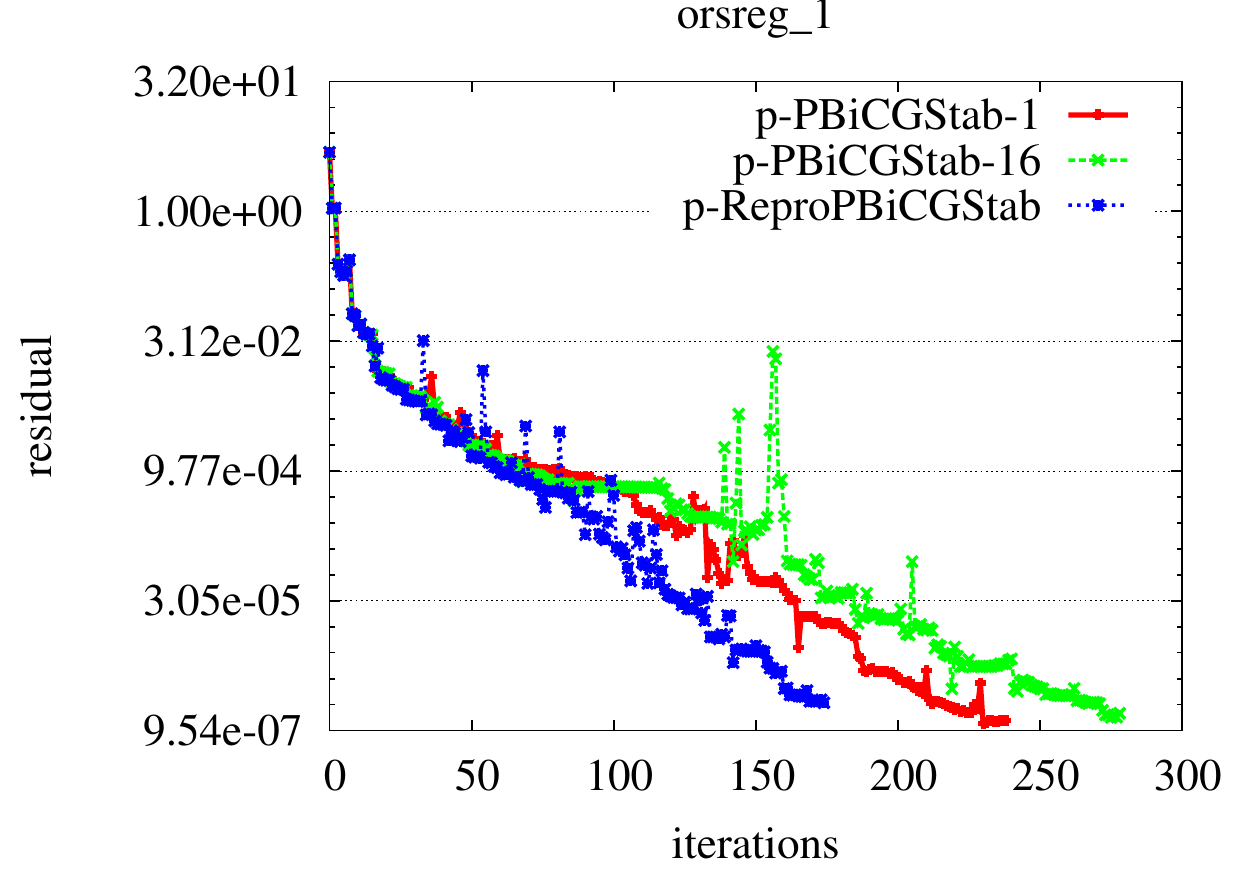}
\end{minipage}
&
\begin{minipage}[t]{0.48\textwidth}
\includegraphics[width=\textwidth,height=6.2cm]{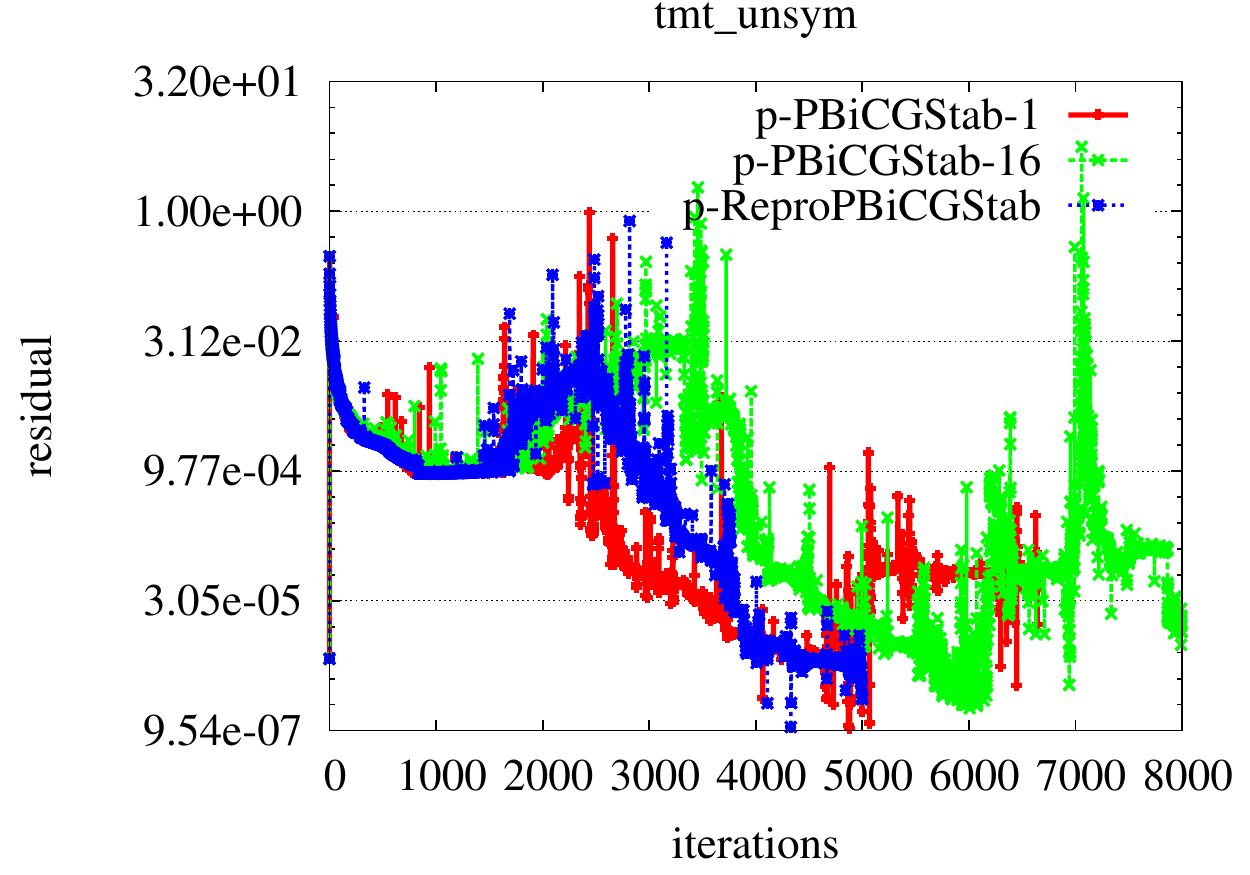}\\
\end{minipage}
\end{tabular}
\vspace{-10pt}
\caption{\label{fig:convergence} Residual history of the standard PBiCGStab and its reproducible variants (first row), and the pipelined PBiCGStab and its reproducible variants (second row); orsreg\_1 results are shown in the first column, while tmt\_unsym in the second column, see~\Cref{tab:ssonenode} for details on matrices. Note that the last iteration is not shown.} 
\end{center}
\end{figure*}

\Cref{fig:convergence} presents the convergence history in terms of the residual computed at every iteration of both the standard and pipelined PBiCGStab methods. The depicted two matrices, namely orsreg\_1 and tmt\_unsym, represent the beneficial scenarios for the reproducible variants, when they reach the approximate solution in significantly less iterations than their non-deterministic variants. In fact, these results demonstrate a sort of idealistic scenario when the reproducible variants converge to the solution faster despite yielding more costly computations per each iteration. Moreover, in these cases, which may not be generic as it fluctuates, the standard and pipelined PBiCGStab require more iterations on 16/ 28 MPI processes.

\begin{table*}[tb]
\begin{center}
\resizebox{\textwidth}{!}{
\begin{tabular}{|l|l|r|r|r|r|r|r|r|r|}
   \hline
   Matrix  & Prec & $N$ & nnz & $\|r^0\|_2$ & \multicolumn{3}{c|}{p-PBiCGStab} & \multicolumn{2}{c|}{p-ReproPBiCGStab} \\  
   \cline{6-8}\cline{9-10}
   & & & & & iter1 & iter16 & $\|b-Ax^j\|_2$ & iter & $\|b-Ax^j\|_2$ \\\hline
   \rowcolor[gray]{.9}add32 & Jac & 4,960 & 19,848 & 8.00e-03 & 36 & 35 & 3.88e-07 & 35 & 6.21e-07\\ 
   af\_shell10 & Jac & 1,508,065 & 52,259,885 & 1.48e+05 & 9 & 9 & 2.18e-02 & 9 & 2.18e-02\\
   \rowcolor[gray]{.9} atmosmodd & Jac & 1,270,432 & 8,814,880 & 3.75e+03 & 222 & 221 & 1.66e-03 & 241 & 3.55e-03\\ 
   atmosmodj & Jac & 1,270,432 & 8,814,880 & 3.75e+03 & 220 & 223 & 2.92e-03 & 223 & 3.53e-03\\ 
   \rowcolor[gray]{.9} atmosmodl & Jac & 1,489,752 & 10,319,760 & 1.85e+04 & 140 & 134 & 1.06e-02 & 132 & 1.82e-02\\ 
   atmosmodm & Jac & 1,489,752 & 10,319,760 & 3.50e+05 & 77 & 78 & 3.37e-01 & 77 & 2.55e-01\\ 
   \rowcolor[gray]{.9} audikw\_1 & Jac & 943,695 & 77,651,847 & 1.58e+07 & 11 & 11 & 7.03e+00 & 11 & 8.04e+00\\   
   bcsstk18 & Jac & 11,948 & 149,090 & 2.30e+09 & 7 & 7 & 7.51e+02 & 7 & 7.51e+02\\
   \rowcolor[gray]{.9}bcsstk26 & Jac & 1,922 & 30,336 & 6.16e+09 & 11 & 11 & 5.62e+03 & 11 & 5.62e+03\\
   bone010 & Jac & 986,703 & 47,851,783 & 8.55e+03 & 12 & 12 & 5.91e-03 & 12 & 5.91e-03\\
   \rowcolor[gray]{.9}boneS10 & Jac & 914,898 & 40,878,708 & 7.17e+03 & 12 & 12 & 3.92e-03 & 12 & 3.92e-03\\   
   Bump\_2911 & Jac & 2,911,419 & 127,729,899 & 1.91e+14 & 12 & 12 & 1.82e+08 & 12 & 1.82e+08\\
   \rowcolor[gray]{.9}cage14 & Jac & 1,505,785 & 27,130,349 & 1.00e+00 & 5 & 5 & 1.55e-07 & 5 & 1.5e-07\\
   cage15 & Jac & 5,154,859 & 99,199,551 & 1.00e+00 & 6 & 6 & 1.12e-07 & 6 & 1.12e-07\\
   \rowcolor[gray]{.9}circuit5M\_dc & Jac & 3,523,317 & 14,865,409 & 1.02e+04 & 5 & 5 & 6.52e-03 & 5 & 6.52e-03\\ 
   CurlCurl\_3 & Jac & 1,219,574 & 13,544,618 & 2.42e+10 & 17 & 17 & 2.14e+04 & 17 & 2.14e+04\\ 
   \rowcolor[gray]{.9} CurlCurl\_4 & Jac & 2,380,515 & 26,515,867 & 2.10e+10 & 19 & 19 & 1.18e+04 & 19 & 1.18e+04\\ 
   ecology1 & Jac & 1,000,000 & 4,996,000 & 1.96e+01 & 9 & 8 & 1.34e-06 & 8 & 9.60e-06\\
   \rowcolor[gray]{.9} ecology2 & Jac & 999,999 & 4,995,991 & 1.96e+01 & 8 & 8 & 7.18e-06 & 9 & 1.07e-05\\ 
   Hardesty1 & Jac & 938,905 & 12,143,314 & 9.99e+00 & 17 & 18 & 5.13e-06 & 20 & 2.13e-06\\ 
   \rowcolor[gray]{.9} ML\_Geer & Jac & 1,504,002 & 110,686,677 & 4.89e+02 & 2426 & 4429 & 5.64e-02 & 3420 & 5.83e-02\\ 
   orsreg\_1 & Jac & 2,205 & 14,133 & 4.83e+00 & 239 & 279 & 1.09e-06 & 175 & 3.03e-06\\ 
   \rowcolor[gray]{.9} Queen\_4147 & Jac & 4,147,110 & 316,548,962 & 1.94e+14 & 52 & 53 & 6.02e+07 & 51 & 5.40e+07\\ 
   rdb3200l & Jac & 3,200 & 18,880 & 9.96e+00 & 671 & 615 & 9.79e-06 & 685 & 8.41e-06\\ 
   \rowcolor[gray]{.9} s3dkq4m2 & Jac & 90,449 & 4,427,725 & 6.08e+02 & 23 & 23 & 7.27e-05 & 23 & 7.27e-05 \\
   tmt\_unsym & Jac & 917,825 & 4,584,801 & 6.45e-06 & 6641 & 7996 & 9.76e-06 & 4999 & 3.01e-10\\ 
   \rowcolor[gray]{.9} Transport & Jac & 1,602,111 & 23,487,281 & 2.45e-02 & 580 & 600 & 5.34e-09 & 582 & 2.32e-08\\
   vas\_stokes\_2M & Jac & 2,146,677 & 65,129,037 & 4.19e-01 & 6880 & 6358 & 2.77e-07 & 6232 & 3.81e-07\\
   \hline
\end{tabular}
}
\end{center}
\caption {\label{tab:ssonenode-pipe-iallreduce-fma} 
Convergence of the {\bf pipelined PBiCGStab and its reproducible versions} (p-ReproPBiCGStab, identical results reported for both) on a set of the SuiteSparse matrices. The initial guess is $x^0 = 0$. The number of iterations required to reach the tolerance of $10^{-6}$ on the scaled residual, i.e. $\|r^j\|_2/\|r^0\|_2$, is reported along with the corresponding true residual $\|b-Ax^j\|_2$.}
\end{table*}







\begin{figure*}[th!]
\begin{center}
\begin{tabular}{cc}
\begin{minipage}[t]{0.48\textwidth}
\includegraphics[width=\textwidth,height=6.2cm]{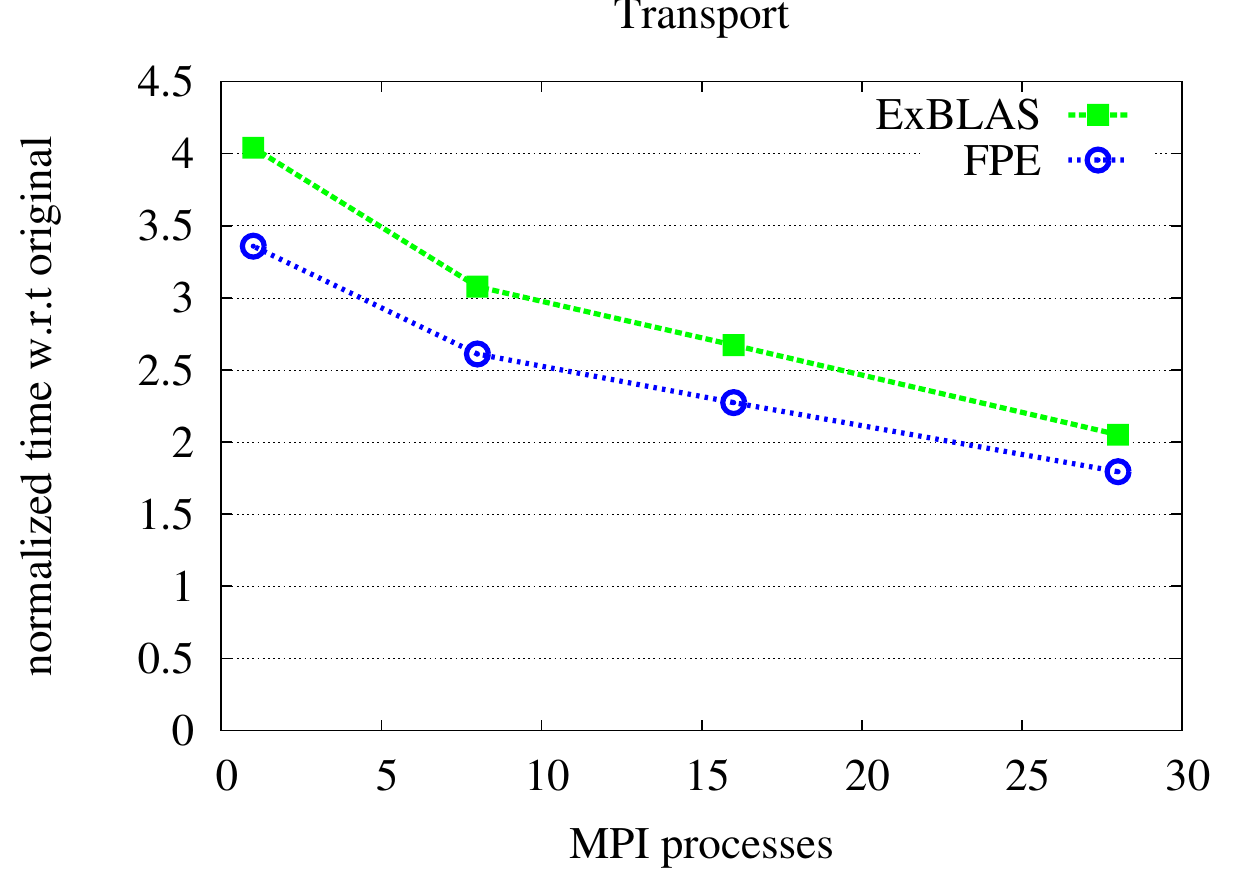}
\end{minipage}
&
\begin{minipage}[t]{0.48\textwidth}
\includegraphics[width=\textwidth,height=6.2cm]{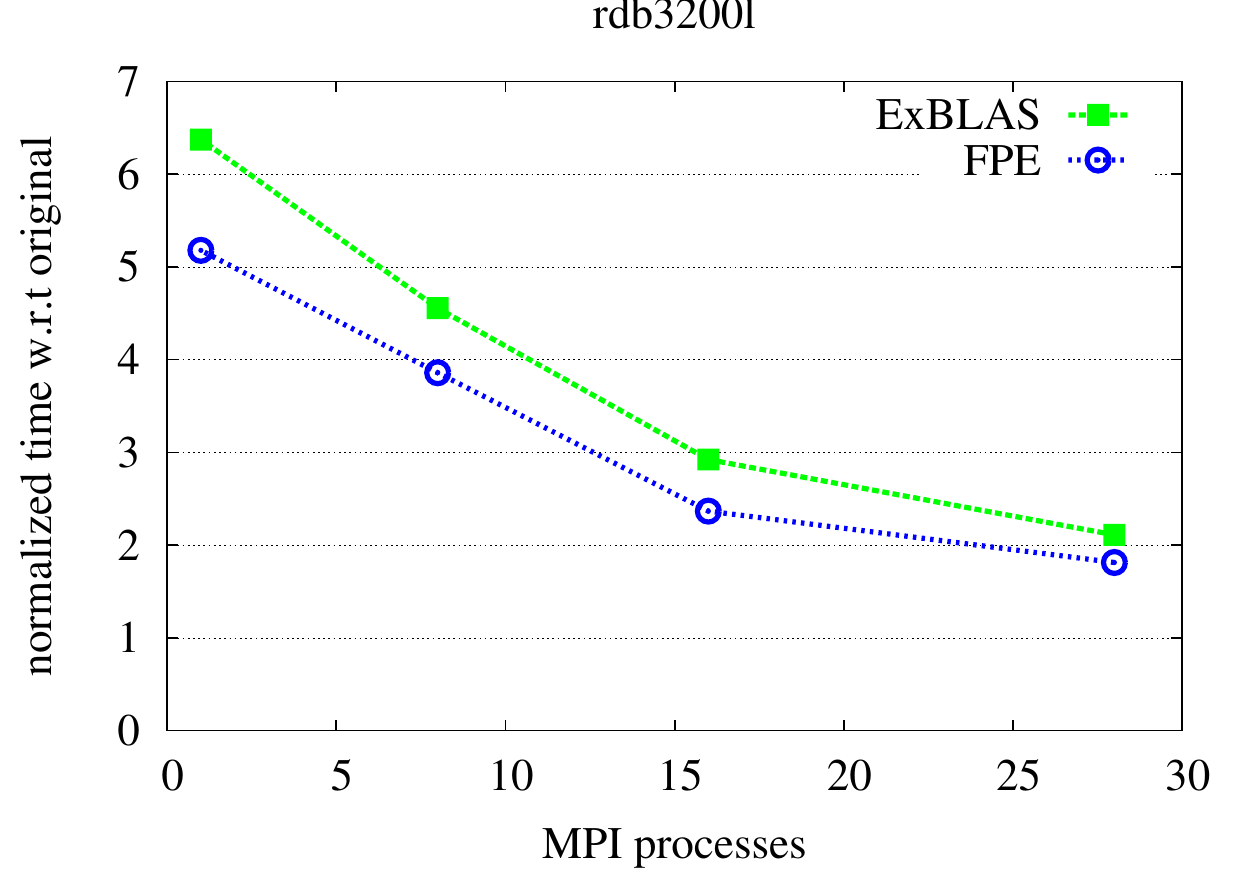}
\end{minipage}
\\
\begin{minipage}[t]{0.48\textwidth}
\includegraphics[width=\textwidth,height=6.2cm]{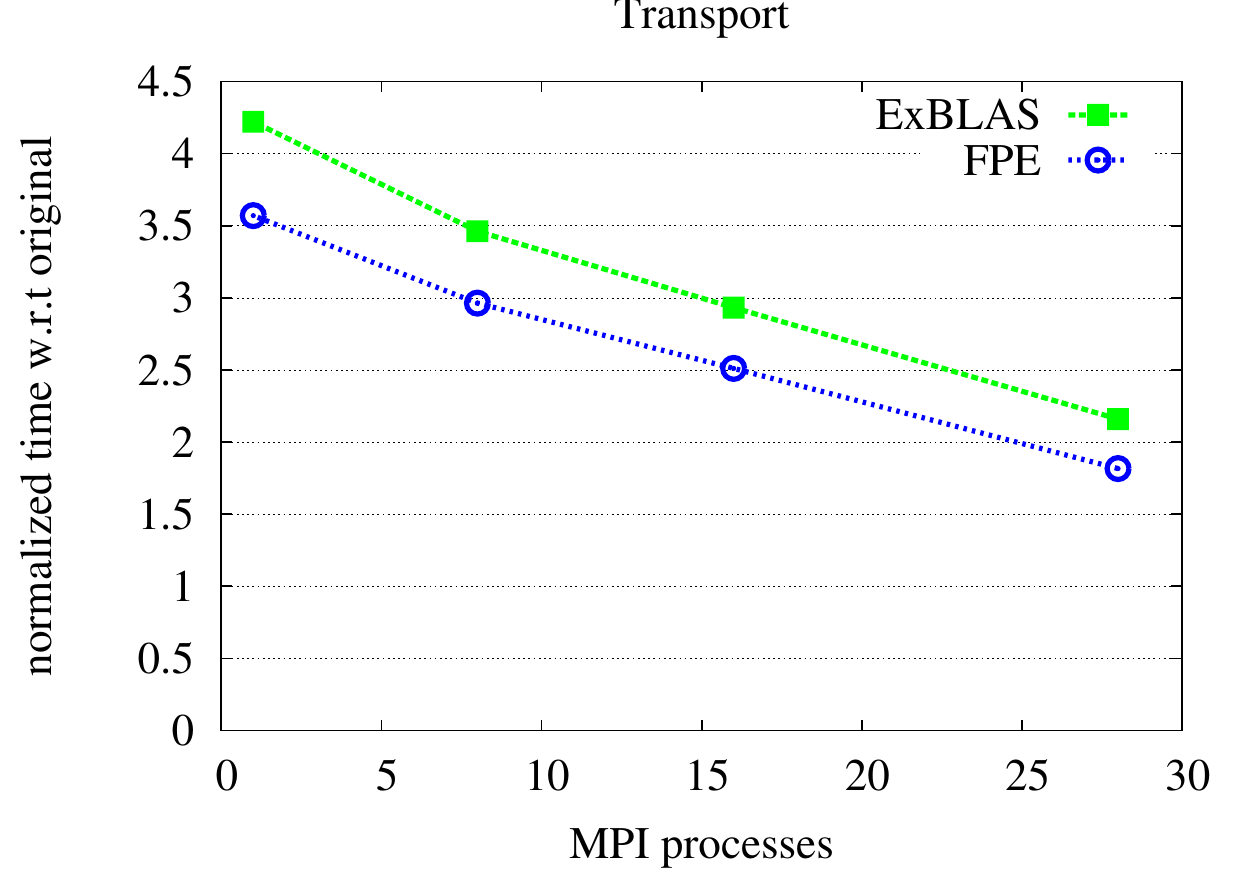}
\end{minipage}
&
\begin{minipage}[t]{0.48\textwidth}
\includegraphics[width=\textwidth,height=6.2cm]{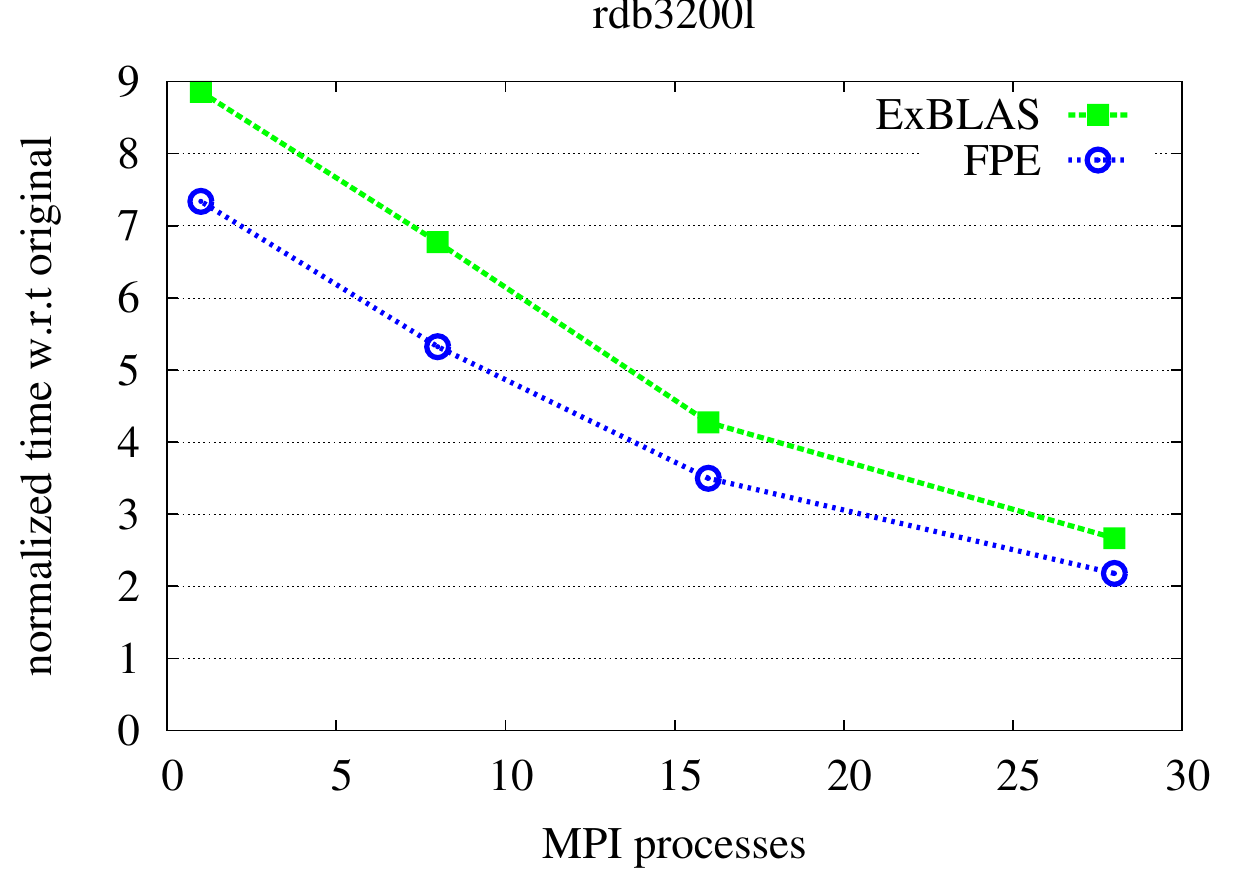}\\
\end{minipage}
\end{tabular}
\vspace{-10pt}
\caption{\label{fig:scale} Strong scaling results of the standard PBiCGStab and its reproducible variants (first row), and the pipelined PBiCGStab and its reproducible variants (second row) on one SkyLake node; Transport results are shown in the first column, while rdb3200l in the second column, see~\Cref{tab:ssonenode} for details on matrices.}
\end{center}
\end{figure*}

\Cref{fig:scale} demonstrates the strong scalability results -- when the problem is fixed but the number of allocated resources varies -- for the original and both \exblas- and FPE-based standard and pipelined PBiCGStab variants on the Transport and rdb3200l matrices. The figure reports the mean execution time for the entire loop of the solver among five samples. We select these matrices due to their variety in the number of nonzero elements: millions for Transport and several thousands for rdb3200l. Note that MPI communication is performed within a node, most likely being exposed to intra-node communication via shared memory. All three variants show good scalability results for Transport with 8.1x (9.0x), 12.0x (12.7x), and 12.4x (13.0x) speed up on 16 MPI processes for the original, FPE, and \exblas\ variants of the standard PBiCGStab (pipelined PBiCGStab), respectively; the corresponding speed up of 4.2x (5.3x) for original, 9.1x (11.0x) for FPE, and 8.6x (10.9x) for \exblas\ runs on rdb3200l. The reproducible variants demonstrate higher speedup due to extra floating-point operations. The overhead of the \exblas\ and FPE variants compared to the original variant is reduced to nearly 2x for both matrices on 28 MPI processes. The scalability on the other matrices from~\Cref{tab:ssonenode,tab:ssonenode-pipe-iallreduce-fma} shows the similar pattern and overhead. However, the smaller number of nonzeros leads to the worse scalability and higher overhead as rdb3200l confirms. For instance, for the orsreg\_1 matrix, the original and ExBLAS/ FPE variants are only 4x and 8x, respectively, faster on 16 MPI processes.

Note that the average execution time per loop for many matrices from~\Cref{tab:ssonenode,tab:ssonenode-pipe-iallreduce-fma} is not sufficient for distributed memory computations. 
This is due to the fact that the potential performance gain from extra nodes is demolished by communication.


\subsection{Scalability} 
 We leverage a sparse s.p.d. coefficient matrix arising from the finite-difference method of a 3D Poisson's equation with 27 stencil points. We perturb the matrix with the values $1.0-0.0001$ below the central point to create the unsymmetric 27-point stencil aka the e-type model~\cite{cools17}. 
The fact that the vector involved in the \spmv kernel has to be replicated in all MPI ranks constrains 
the size of the largest problem that can be solved. 
Given that the theoretical cost of PBiCGStab is $t_c \approx 4nnz+ 26n$ floating-point arithmetic operations, 
where $nnz$ denotes  the number of nonzeros of the original matrix and its size $n$, 
the execution time of the method is usually dominated by that of the \spmv kernel.
Therefore, in order to analyze the weak scalability of the method, we maintain the number of non-zero entries per node. 
For this purpose, 
we modified the original matrix, transforming it into a band matrix, 
where the lower and upper bandwidths (\textit{bandL} and \textit{bandU}, respectively) 
depend on the number of nodes employed in the experiment as follows:
\vspace*{-2mm}
$$
\vspace*{-2mm}
bandL = bandU = 100 \times \#nodes \quad \rightarrow 
$$
$$
nnz = ( bandL + bandU + 1 ) \times n.
$$
With 32 nodes, the bandwidth ranges between 100 and 3200. 
With this approach we can then maintain the number of rows/ columns of the matrix equal to $n$=4M (4,019,679), while increasing its bandwidth and, therefore, the computational workload
proportionally to the hardware resources, as required
in a weak scaling experiment. 

The right-hand side vector $b$ in the iterative solvers was always initialized to the product of $A$ with a vector
containing ones only; 
and the PBiCGStab iteration was started with the initial guess $x_0=0$. 
The parameter that controls the convergence of the iterative process was set to $10^{-6}$.

\begin{figure*}[tb]
\begin{center}
\begin{tabular}{cc}
\begin{minipage}[t]{0.49\textwidth}
\hspace*{-4mm}\includegraphics[width=1.05\textwidth]{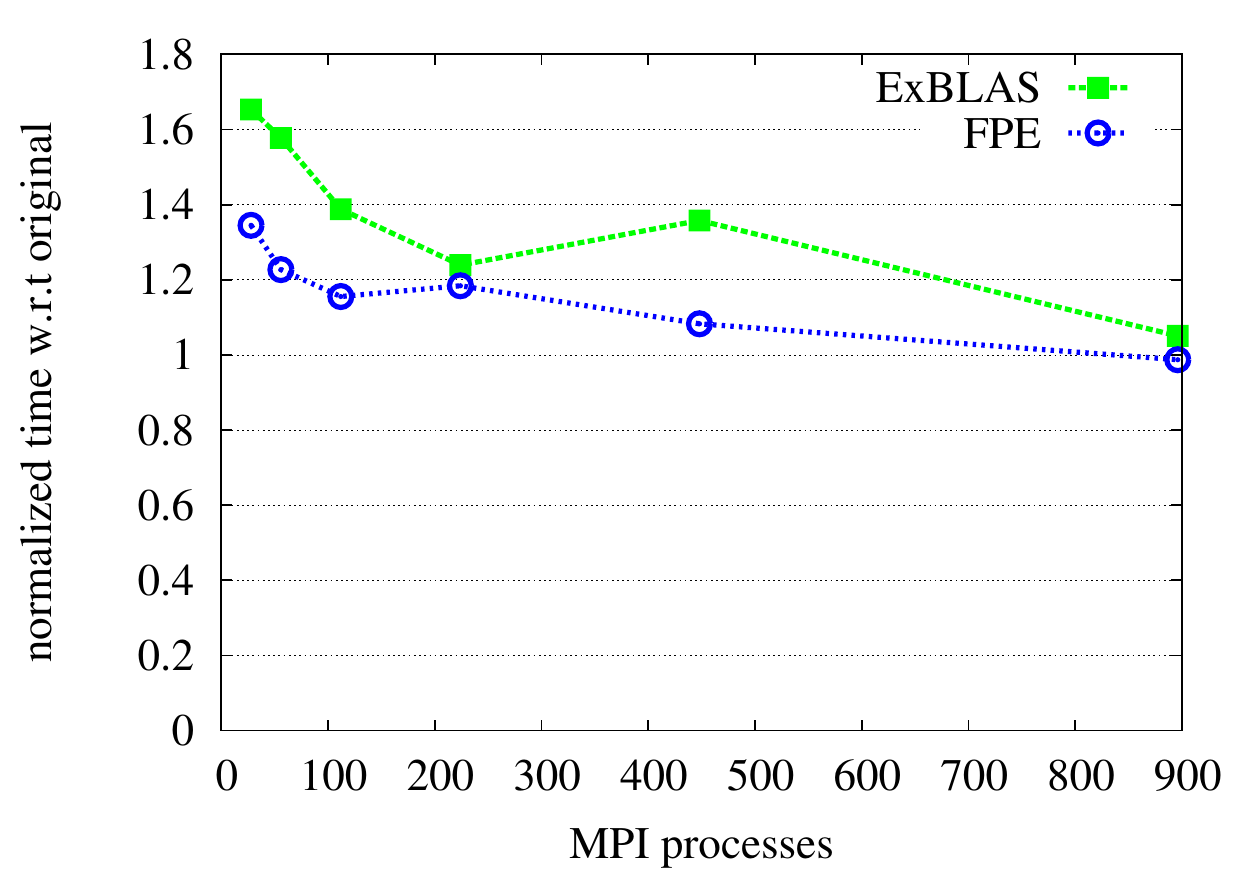}
\end{minipage}
&
\begin{minipage}[t]{0.49\textwidth}
\hspace*{-4mm}\includegraphics[width=1.05\textwidth]{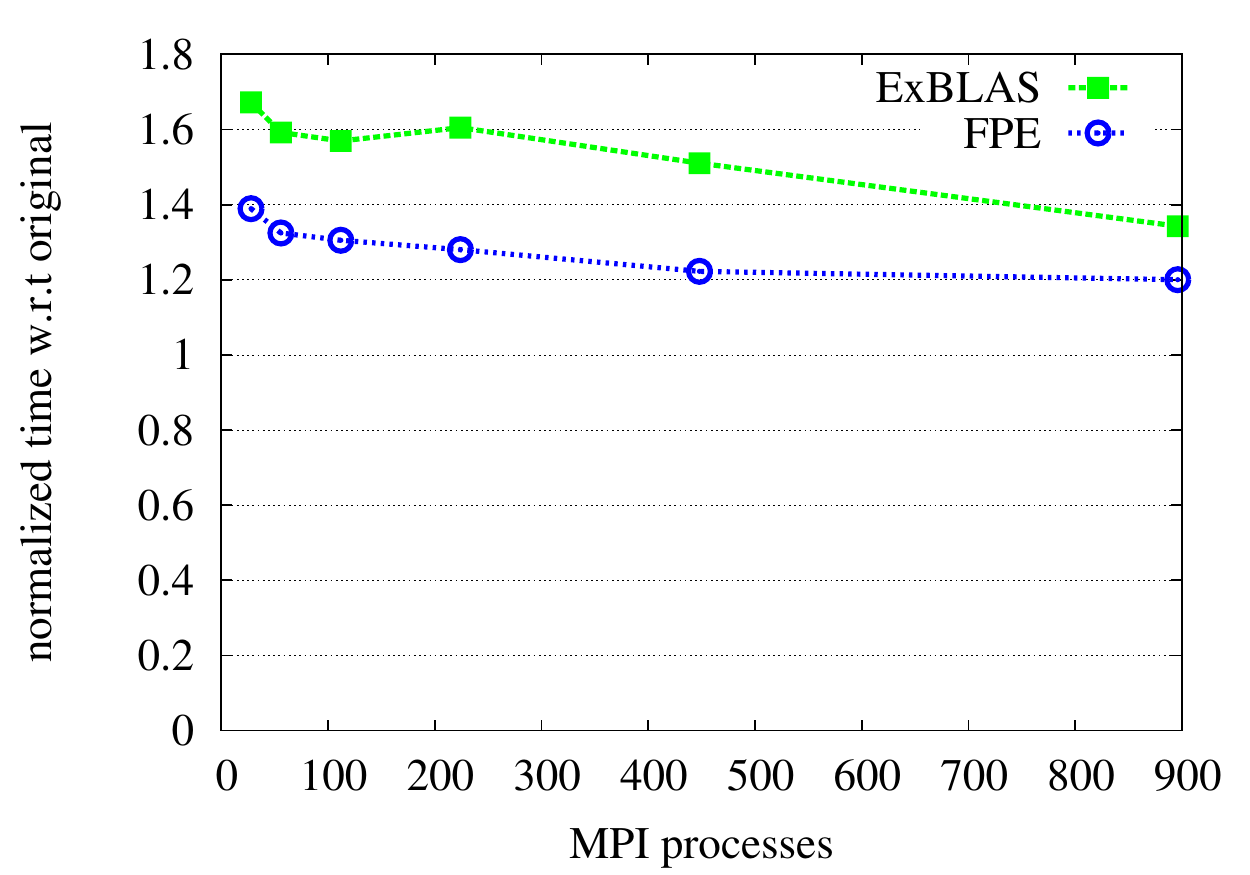}
\end{minipage}
\end{tabular}
\vspace{-5pt}
\caption{\label{fig:scaling} Strong (left) and weak (right) scalability of the reproducible PBiCGStab variants with the normalized time against the non-deterministic MPI variant.}
\end{center}
\end{figure*}

\Cref{fig:scaling} reports the results of both strong and weak scaling for the reproducible variants against the original version. For the strong scaling, we fix the problem to 16M non-zeros and varied the number of nodes/ cores used, while for the weak scaling the work load per node is kept constant as 4M non-zeros by varying the bandwidth with respect to the number of nodes involved; presumably, there is enough load to hide the impact of communication. 
For both scalability cases, the initial overhead is the same, namely 67\,\% for the version with ExBLAS and 38-40\,\% for FPE. With the strong scaling, the overhead reduces to 8.2\,\% for ExBLAS and 3.0\,\% for FPE as the communication starts to take over and the overhead between the two versions narrows. However, this pattern is not preserved for the weak scaling where the ExBLAS version of PBiCGStab shows about 38\,\% overhead on 32 nodes. 

\subsection{Accuracy and Reproducibility} In addition, we derive a sequential version of the PBiCGStab as in~\Cref{fig:pbicgstab} that relies on the GNU Multiple Precision Floating-Point Reliably (MPFR) library~\cite{MPFR} -- a C library for multiple (arbitrary) precision floating-point computations on CPUs -- as a highly accurate reference implementation. This implementation uses 2,048 bits of accuracy for computing \dotp product, 192 bits for internal element-wise product, and performs correct rounding of the computed result to double precision. 

\Cref{tab:acc} reports the intermediate and final (except from original that takes longer) scaled residual on each iteration of the PBiCGStab solvers for the orsreg\_1 matrix, as in~\Cref{tab:ssonenode}, under the tolerance of $10^{-6}$ on eight MPI processes. We also add the results of the original code on one core/ process to highlight the reproducibility issue. The results are presented with all digits using hexadecimal representation. We report only few iterations, however the difference is present on all iterations. The sequential MPFR version confirms the accuracy and reproducibility of parallel \exblas\ and FPE variants by reporting identical number of iterations, intermediate residuals, and both the final true and initial scaled residuals. However, the MPFR variant of PBiCGStab converges to the approximate solution in 3.39e-01 seconds, while the \exblas\ and FPE variants take 3.95e-02 and 2.75e-02 seconds (8.57x and 12.32x faster), accordingly, on eight MPI processes.
The original code shows the discrepancy from few digits on the initial iteration and up to almost the entire number on the final iterations; the count of required iterations also differs from the reproducible and MPFR variants.

\begin{table*}[tb]
\begin{center}
\resizebox{\textwidth}{!}{%
\begin{tabular}{|c|l|l|l|l|}
   \hline
   Iteration  & \multicolumn{4}{c|}{Residual} \\ 
   \cline{2-5}
   & {\em MPFR} & {\em Original} 1 proc & {\em Original} 8 procs & {\em Exblas \& FPE} \\ \hline  \hline
   \rowcolor[gray]{.9} 0 & 0x1.3566ea57eaf3fp+2 & 0x1.3566ea57ea{\bf b49}p+2 & 0x1.3566ea57ea{\bf b49}p+2 & 0x1.3566ea57eaf3fp+2\\
   1 & 0x1.146d37f18fbd9p+0 & 0x1.146d37f18f{\bf aaf}p+0 & 0x1.146d37f18f{\bf ab}p+0 & 0x1.146d37f18fbd9p+0\\
   \rowcolor[gray]{.9} ... & ... & ... & ...& ...\\
   99 & 0x1.cedf0ff322158p-13 & {\bf 0x1.88008701ba87p-12} & {\bf 0x1.04e23203fa6fcp-12} & 0x1.cedf0ff322158p-13\\
   \rowcolor[gray]{.9}100 & 0x1.be3698f1968cdp-13 & {\bf 0x1.55418acf1af27p-12} & 0x1.{\bf fbf5d3a5d1e49}p-13 & 0x1.be3698f1968cdp-13\\
    ... & ... & ... & ...& ...\\
   \rowcolor[gray]{.9}208 & 0x1.355b0f18f5ac1p-20 & {\bf 0x1.19edf2c932ab8p-18} & 0x1.{\bf b051edae310c7}p-20 & 0x1.355b0f18f5ac1p-20\\
   209 & 0x1.114dc7c9b6d38p-20 & {\bf 0x1.19b74e383f74ep-18} & 0x1.{\bf a18fc929018d4}p-20 & 0x1.114dc7c9b6d38p-20\\
   \rowcolor[gray]{.9}210 & 0x1.03b1920a49a7ap-20 & {\bf 0x1.19c846848f361p-18} & 0x1.{\bf c7eb5bbc198b1}p-20 & 0x1.03b1920a49a7ap-20\\
   \hline
\end{tabular}
}
\end{center}
\caption {\label{tab:acc} Accuracy and reproducibility of the intermediate and final residual against MPFR for the orsreg\_1 matrix, see~\Cref{tab:ssonenode}. } 
\end{table*}

\section{Related Work}
\label{sec:related:works}

To enhance reproducibility, Intel proposed the ``Conditional Numerical Reproducibility'' (CNR) option in its Math Kernel Library (MKL).
Although CNR guarantees reproducibility, it does not ensure correct rounding, meaning the accuracy is arguable. 
Additionally, the cost of obtaining reproducible results with CNR is high. For instance, for large arrays 
the MKL's summation with CNR was almost 2x slower than the regular MKL's summation on the Mesu cluster hosted at the Sorbonne University~\cite{Collange15Parco}.

Demmel and Nguyen implemented a family of algorithms -- that originate from the works by Rump, Ogita, and Oishi~\cite{RuOgOi2010,RuOgOi2008b} -- for reproducible summation in floating-point arithmetic~\cite{Demmel13Arith0,Demmel14OneRed}. These algorithms always return the same answer. They first compute an absolute bound of the sum and then round all numbers to a fraction of this bound. In consequence, the addition of the rounded quantities is exact, however the computed sum using their implementations with two or three bins is not correctly rounded. 
Their results yielded roughly $20$\,\% overhead on $1024$ processors (CPUs only) compared to the Intel MKL \texttt{dasum()}, but it shows $3.4$ times 
slowdown on $32$ processors (one node). 
Ahrens, Nguyen, and Demmel extended their concept to few other reproducible BLAS routines, distributed as the ReproBLAS 
library (\url{http://bebop.cs.berkeley.edu/reproblas/}), but only with parallel reproducible reduction. Furthermore, the ReproBLAS effort was extended to reproducible  tall-skinny QR~\cite{NguyenD15}.

The other approach to ensure reproducibility is called ExBLAS, which is initially proposed by Collange, Defour, Graillat, and Iakymchuk in~\cite{Collange15Parco}. ExBLAS is based on combining long accumulators and floating-point expansions in conjunction with error-free transformations. This approach is presented in~\Cref{sec:background}. Collange et al. showed~\cite{Collange15Parco} that their algorithms for reproducible and accurate summation have $8$\,\% overhead on $512$ cores (32 nodes) and less than $2$\,\% overhead on 16 cores (one node). While ExSUM covers wide range of architectures as well as distributed-memory clusters, the other routines primarily target GPUs. Exploiting the modular and hierarchical structure of linear algebra algorithms, the ExBLAS approach was applied to construct reproducible LU factorizations with partial pivoting~\cite{Iakymchuk19ReproLU}.

Mukunoki and Ogita presented their approach to implement reproducible BLAS, called OzBLAS~\cite{ozblas}, with tunable accuracy. This approach is different from both ReproBLAS and ExBLAS as it does not require to implement every BLAS routine from scratch but relies on high-performance (vendor) implementations. Hence, OzBLAS implements the Ozaki scheme~\cite{Ozaki2012} that follows the fork-join approach: the matrix and vector are split (each element is sliced) into sub-matrices and sub-vectors for secure products without overflows; then, the high-performance BLAS is called on each of these splits; finally, the results are merged back using, for instance, the NearSum algorithm. Currently, the OzBLAS library includes dot product, matrix-vector product (gemv), and matrix-matrix multiplication (gemm). These algorithmic variants and their implementations on GPUs and CPUs (only dot) reassure reproducibility of the BLAS kernels as well as make the accuracy tunable up-to correctly rounded results. 

The proposed framework was implicitly used to derive the reproducible preconditioned Conjugate Gradient (PCG) variants with the flat MPI~\cite{iakymchuk19jcam} and hybrid MPI plus OpenMP tasks~\cite{iakymchuk20ijhpca}. The reproducible PCG variants were primarily verified on the 3D Poisson's equation with 27 stencil points showing the good scalability and low performance overhead (under 30\,\% for both the \exblas\ and lightweight variants) on up to 768 cores of the MareNostrum4 cluster.

\section{Conclusions}
\label{sec:conclusion}
Parallel Krylov subspace methods may exhibit the lack of reproducibility when implemented in parallel environments  as the results in~\Cref{tab:ssonenode,tab:ssonenode-pipe-iallreduce-fma,tab:acc} confirm. Such numerical reliability is needed for debugging and validation~\&~verification. In this work, we proposed a general framework for re-constructing  reproducibility and re-assuring accuracy in any Krylov subspace method. Our framework is based on two steps: analysis of the underlying algorithm for numerical abnormalities; addressing them via algorithmic solutions and programmability hints. The algorithmic solutions are build around the \exblas\ project, namely: \exblas\ that effectively combines long accumulator and FPEs; FPEs for the lightweight version. 
The programmability effort was focused on: explicitly invoking \fma\ instructions to avoid replacements by compilers; customized and \fma-based \axpy\ and \axpy-like operations instead of MKL or similar BLAS libraries; as well as to postpone the division to the moment where it is required. 

As test cases, we used the preconditioned standard and pipelined BiCGStab methods and derived two reproducible algorithmic variants for each of them. It is worth mentioning that the two BiCGStab methods are in fact different algorithms with different set of operations yielding non-identical computation path and the divergent way rounding errors are propagate; this difference can be witnessed by the convergence history in~\Cref{fig:convergence} even when using the reproducible variants. The reproducible variants deliver identical results of the standard and also pipelined PBiCGStab, which are confirmed by its MPFR version, to ensure reproducibility in the number of iterations, the intermediate and final residuals, as well as the sought-after solution vector. 
We verified our implementations on a set of the SuiteSparse matrices, showing the performance overhead of nearly 2.0x for the \exblas\ and FPE-based versions, with a noticeably lower overhead for the latter. 
Tests with the 27-point stencil on 32 nodes show almost negligible overhead of 8\,\% and 3\,\%, respectively. 
The code is available at \url{https://github.com/riakymch/ReproPBiCGStab}.

With this study we want to promote reproducibility by design through the proper choice of the underlying libraries as well as the careful programmability effort. 
For instance, a brief guidance would be 1) for fundamental numerical computations use reproducible underlying libraries such as ExBLAS, ReproBLAS, or OzBLAS~\cite{ozblas}; 2) analyze the algorithm and make it reproducible by eliminating any uncertainties and non-deterministic order of computations that may violate associativity such as reductions and use/ non-use of {\fma} and postponing divisions until actually needed. 
Additionally, we try to argue the need for the bit-wise reproducible and correctly-rounded results for iterative solvers as they will anyway be enhanced on next iterations as we do not reach the desired tolerance and, thus, do not exploit at full the obtained bit-wise results. This becomes more evident with the mixed-precision approaches, which we foresee to pursue. 

Our future work is to investigate the residual replacement strategy in the pipelined Krylov subspace solvers such as the pipelined PBiCGStab (p-PBiCGStab)~\cite{cools17} and to study if such strategy can be mitigated by the higher precision provided by long accumulator and FPEs. 
We believe that there is a potential of using higher precision provided by long accumulator and FPEs in order to mitigate the different way rounding errors are propagate as well as to cope with the attainable precision loss in p-PBiCGStab.  

\begin{acks}
This research was partially supported by the EU H2020 MSCA-IF Robust project (No. 842528); the EU H2020 CoE CEEC (No. 101093393); the French ANR InterFLOP project (No. ANR-20-CE46-0009). The research from Universitat Jaume I was funded by the project PID2020-113656RB-C21 via MCIN/AEI/10.13039/501100011033 and project UJI-B2021-58.
\end{acks}

\bibliographystyle{SageH}

\end{document}